\documentclass[a4paper,12pt]{article}

\usepackage[left=2cm,right=2cm,top=2cm,bottom=3cm]{geometry}
\usepackage{amsthm,amsmath}
\usepackage{graphicx}
\usepackage[utf8]{inputenc}

\usepackage{amssymb}

\usepackage[colorlinks,allcolors=blue]{hyperref}

\providecommand{\href}[2]{\texttt{#2}}
\providecommand{\url}[1]{\texttt{#1}}

\newcommand{\tsedoi}[1]{\href{http://doi.org/#1}{\texttt{#1}}}
\newcommand{\arXivnumber}[1]{\href{http://arxiv.org/abs/#1}{\texttt{arXiv:#1}}}

\newcommand{\vvset}[1]{{\mathcal{#1}}}
\newcommand{\Acal}{\vvset{A}}
\newcommand{\Bcal}{\vvset{B}}

\newcommand{\Ecal}{\vvset{E}}
\newcommand{\Fcal}{\vvset{F}}
\newcommand{\Gcal}{\vvset{G}}
\newcommand{\Hcal}{\vvset{H}}

\newcommand{\Ncal}{\vvset{N}}

\newcommand{\Vcal}{\vvset{V}}

\newcommand{\vvmatr}[1]{\mathbf{#1}}
\newcommand{\trans}{\mathrm{T}} 
\newcommand{\Amatr}{\vvmatr{A}}
\newcommand{\Amatrtilde}{\tilde{\Amatr}}
\newcommand{\Gtilde}{\tilde{G}}
\newcommand{\Shat}{\hat{S}}
\newcommand{\Wmatr}{\vvmatr{W}}

\newcommand{\Afrak}{\mathfrak{A}}
\newcommand{\Afraktrivial}{\Afrak_\mathrm{trivial}}

\newcommand{\Ffrak}{\mathfrak{F}}

\newcommand{\Ebb}{\mathbb{E}}

\newcommand{\simfunc}{\mathrm{sim}}
\newcommand{\simfuncnull}{\mathrm{sim}_\mathrm{null}}
\newcommand{\Ncalpre}{\Ncal^{\mathrm{(pre)}}}
\newcommand{\Ncalsuc}{\Ncal^{\mathrm{(suc)}}}

\newcommand{\kin}{k^{(\mathrm{in})}}
\newcommand{\kout}{k^{\mathrm{out}}}

\newcommand{\texpect}[1]{\langle #1 \rangle}

\newcommand{\bea}{\begin{eqnarray}}
\newcommand{\eea}{\end{eqnarray}}
\newcommand{\beq}{\begin{equation}}
\newcommand{\eeq}{\end{equation}}

\providecommand{\eqref}[1]{(\ref{#1})}
\newcommand{\figref}[1]{Fig.\ \ref{#1}}
\newcommand{\Figref}[1]{Fig.\ \ref{#1}} 
\newcommand{\secref}[1]{section \ref{#1}}
\newcommand{\appref}[1]{Appendix} 
\newcommand{\appsecref}[1]{Appendix}
\newcommand{\tabref}[1]{Table \ref{#1}}

\newcommand{\tdef}[1]{\textsc{#1}}

\newcommand{\tpre}[1]{}

\newcommand{\tnote}[1]{} 
\newcommand{\tcomment}[1]{} 
\newcommand{\vnote}[1]{} 
\newcommand{\vcomment}[1]{} 

\begin{document}

\vspace*{0.2in}

\renewcommand{\thefootnote}{\fnsymbol{footnote}}

\begin{center}
{\Large\textbf{Making Communities Show Respect for Order\footnote{Version for \texttt{arXiv} of paper published 
  as Applied Network Science \textbf{5} (2020) 15,
  DOI: \href{http://doi.org/10.1007/s41109-020-00255-5}{\texttt{http://doi.org/10.1007/s41109-020-00255-5}},
  \href{http://arxiv.org/abs/1908.11818}{\texttt{arXiv:1908.11818}}. }
}} \\[6pt]
 {\large {V.\ Vasiliauskaite}\footnote{Corresponding Author.}},
 {\large {T.S.\ Evans}}
 \\[6pt]
Centre for Complexity Science and  Theoretical Physics Group,
\\ Imperial College London, SW7 2AZ, U.K.
\\ 21st February 2020
\end{center}
\begin{abstract} 
In this work we give a community detection algorithm in which the communities both respects the intrinsic order of a directed acyclic graph and also finds similar nodes. We take inspiration from classic similarity measures of bibliometrics, used to assess how similar two publications are, based on their relative citation patterns. We study the algorithm's performance and antichain properties in artificial models and in real networks, such as citation graphs and food webs. We show how well this partitioning algorithm distinguishes and groups together nodes of the same origin (in a citation network, the origin is a topic or a research field). We make the comparison between our partitioning algorithm and standard hierarchical layering tools as well as community detection methods. We show that our algorithm produces different communities from standard layering algorithms.
\end{abstract}

\setcounter{footnote}{0}
\renewcommand{\thefootnote}{\arabic{footnote}}

\section{Introduction}

Nodes in networks have many natural orders. Every centrality measure allows us to say if one node has a higher centrality value than another. However, some systems can have important constraint leading to a characteristic order in a network; examples include the publication dates of papers in a citation network, dependency of packages in computer software, and predator-prey relationships in a food web. If edges respect this order, they exist only if they link a high value node to a lower value node, from an earlier paper to a later paper, then edges are directed and there can be no cycles --- a Directed Acyclic Graph (DAG).

A link between two nodes in a DAG must always encode the \textit{order} of the pair.  It is the converse which is important here: certain links can not exist because of the inherent order.  This means that two nodes can be very similar but they can not be connected while in standard network analysis such an edge would be assumed to exist most of the time in a DAG.
For example, two papers can be produced independently at the same time with similar new results yet by definition they can not cite each other.
The development of the relativistic model for the Higgs mechanism is a good illustration of this as it is attributed to three independent groups: Brout and Englert (August 1964), Higgs (October 1964), and  Guralnik, Hagen and Kibble (November 1964). Only the last of these three papers cites the two earlier publications but there is no citation between the first two.

So in order to understand DAGs, we need to adapt our standard network analysis tools to make them respect the order implicit in these common network topologies \cite{KN09,CE14,C14a,CE16}.
In this paper we will focus on the important topic of community detection in networks, clustering in the language of data science, in which the aim is to find sets of similar nodes~\cite{F09,JM99}. Since most network methods assume that a link between nodes indicates similarity,
network communities translate the similarity of nodes into the requirement that communities are subgraphs where there are more links within the community than there are to the rest of the network~\cite{F09}. This is justified when links indicate node similarity that is not necessarily true for DAGs so we need to find new ways to define clusters otherwise the order in a DAG may obscure some of the natural clustering.

\begin{figure}
\centering
\includegraphics[width = 0.6\textwidth]{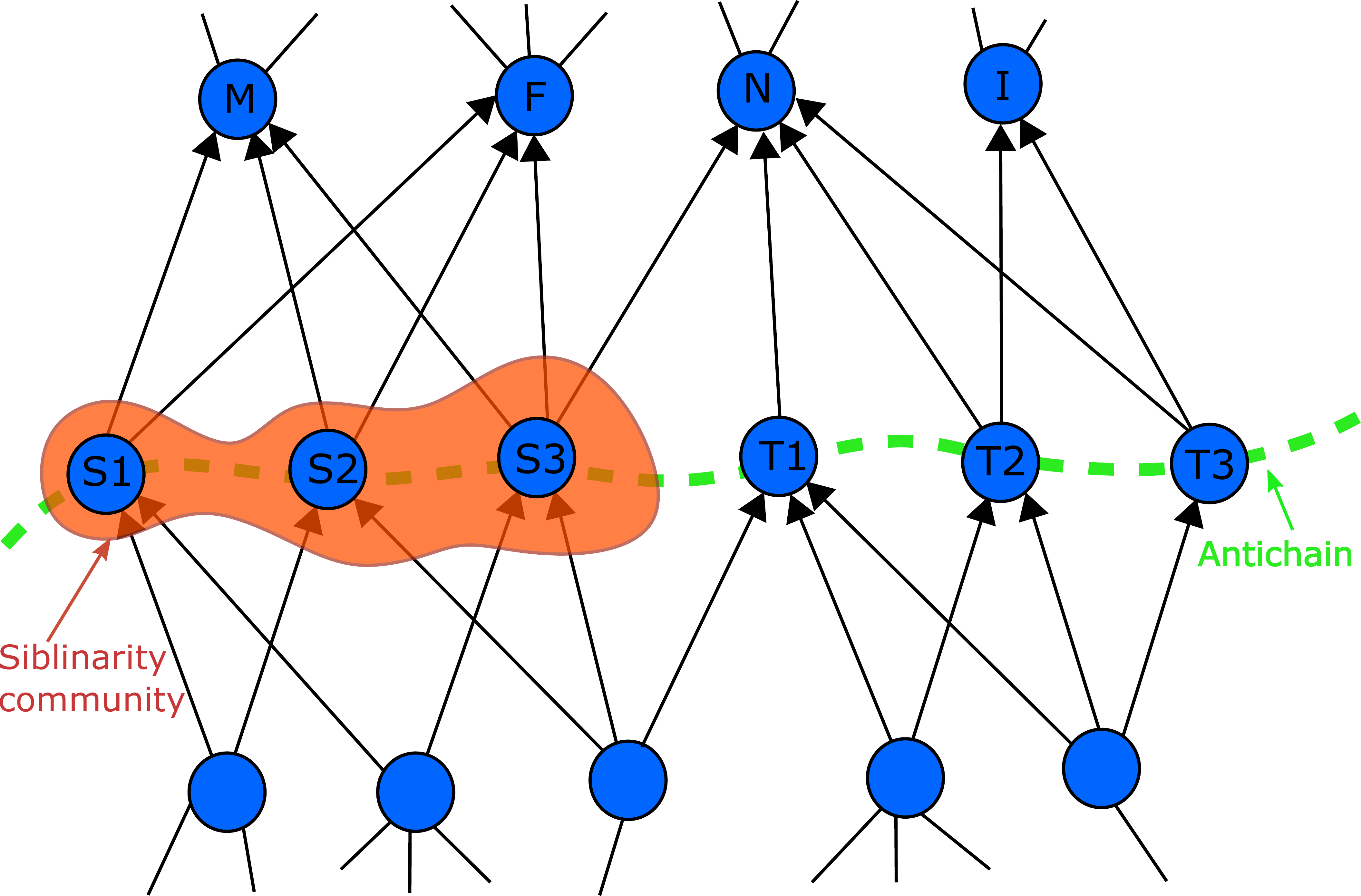}
\caption{An antichain is a subset of nodes in the graph, such that none of the nodes are pairwise connected with edges or paths. In this illustration, an antichain is composed of nodes $\{S_1,S_2,S_3,T_1,T_2,T_3\}$ (the green dotted line is drawn as a guide of eye). Although in the same antichain, not all nodes share many neighbours. Nodes $\{S_1,S_2,S_3\}$ share many successors (they are all connected to nodes $M$, $F$), so we join them in a Siblinarity community.
}
\label{ftoyexample}
\end{figure}

To see how we will find communities in a DAG, we start by noting that one expression of the order in a DAG is that there is a natural hierarchy in the system.
Two computer packages which fill a very similar role will not depend on each other but they will draw on similar ``lower level'' packages reflecting a hierarchy and an order in packages. For instance, the python network package  \texttt{networkx} is a prerequisite for two python community detection algorithms, \texttt{python-louvain} and \texttt{demon}.
In the Florida-bay foodweb we study in Section \ref{sfoodweb}, pinfish, parrotfish and manatee are grouped together because they feed on similar species, such as detritivorous polychaetes, sponges, bivalves. So in a DAG a natural property of any nodes at the same level in the hierarchy is that they are \emph{not} connected, directly by an edge or even indirectly via a longer path. So, in the context of a network with an order, with a hierarchy, it is very natural to cluster nodes which are not connected by any path, and these are called \tdef{antichains}, see \figref{ftoyexample}. So in our approach, in order to respect the order inherent in a DAG we will create communities which are antichains.

The next problem is that there are many possible antichains and each node can be in many different antichains. Also nodes in useful communities will need to be similar in some sense. In a set of computer packages, we don't want to cluster a package on games with one on networks, rather we want to collect all the different network packages together in one community.  Likewise, we don't want to cluster species at the same level in food webs if they are from completely different environments. So we will aim to find the antichains which contain nodes which are similar by some appropriate measure.
We take our inspiration from classic measures used to assess the similarity of two documents from their citation network alone (for example see \cite{BK10a} for more recent application).
In \textit{bibliographic coupling} the similarity of two documents is measured using the overlap of their bibliographies~\cite{K63,M64}.  The \textit{co-citation} similarity of two documents~\cite{S73,SG74} uses the overlap in the citing documents. So by looking into similarities of neighbourhoods of two nodes, we can say something about how similar they are themselves. Though defined in terms of a citation network, these similarity measures are widely used in network analysis, for example see \cite{SP11}.

Our aim in this paper is to produce a method to find communities of similar nodes which take account of the sense of order in a DAG, the hierarchy implicit in a DAG.  We  partition the nodes of a DAG into antichains which have large neighbourhood overlaps.  We will also look at the properties of our DAG clusters and we will compare them against properties of other types of DAG layering and clustering methods. Lastly we will investigate the potential uses of this approach to study citation networks and food webs.


Our method is related to two different types of network algorithms: community detection and DAG layering. Community detection does not include a notion of ``layers'' but aims to find a collection of nodes with strong intra-community links.  In this sense a community of similar nodes is being defined by the network topology on mesoscopic scales, not just on the local scale of nearest neighbours. Since an edge connecting two nodes is normally used as the strongest indicator of a relationship, antichains seem at first to be the exact opposite of traditional network communities. As a consequence, a conventional community detection method applied to a DAG (especially if the edge direction is ignored) can be composed of nodes from many different layers. DAG layering, on the other hand, does not have a notion of similarity but typically it aims to minimise the number of layers used so maximising the size of the layers.  Such layers are composed of hierarchically equivalent nodes but a layer does not represent a community of similar nodes. Our algorithm bridges this gap and finds layer decomposition of a DAG with a notion of similarity.

Community detection is an active research area in network science. This field is wide and many different approaches to finding groups of similar nodes were proposed. An extensive discussion of the main approaches to community detection in networks is given in~\cite{F09}.
Some works have studied the stability and performance of conventional community detection algorithms in DAGs such as~\cite{LCSN07,BKZ10}. There is also some work on specialised approaches to community detection in directed acyclic graphs, for instance~\cite{STM15}, in addition to clustering algorithms developed for analysis of task scheduling graphs (which are also DAGs), for example see~\cite{GY92} and references therein.

Graph layering is another topic related to our work. It is used for drawing hierarchical graphs, such as trees and DAGs. Layering is a problem of partitioning a DAG into layers, such that edges in the visualisation always point in one direction on the page, thus it is a form of partitioning a DAG into antichains. As part of Sugyiama graph drawing algorithm, various layering techniques are used to remove edge overlap as much as possible, as well as minimise the number of edges that span many layers~\cite{STT81}. Perhaps the simplest way to partition a DAG into layers (or antichains) is by using the so-called longest path algorithm, which yields the smallest number of layers~\cite{M71}. We will discuss this algorithm in more detail in \secref{sheightpartition} and utilise partitions, obtained using the longest paths, to evaluate the performance of our proposed method. More sophisticated graph layering algorithms limit the number of layers in the graph, so the size of most layers is large, for instance to minimise the number of layers with crossing edges~\cite{TH13,GKNV93,NT06} or to ensure the maximum width of any of the layers~\cite{HN02,HN13}.

Our aim here is to find a clustering algorithm which both respects the topology and ordering of a DAG while using that same topology to ensure the clusters contain only similar nodes.
We start by discussing the methods we use in \secref{smethods}. We look at the relevant properties of DAGs in \secref{sdagproperties}.  We review a standard way to partition a DAG into antichains by considering the heights and depths of nodes in \secref{sheightpartition}.
We then describe the algorithm we used to find our siblinarity communities in \secref{ssiblinaritypartiton} and \secref{soptimisation}.
In \secref{datamodels} we describe the models and data used to test our algorithm.
The results of our analysis are given in \secref{sresults} where we study the properties of siblinarity-based communities, comparing them to alternative methods and exploring variations of our algorithm. We conclude with a brief overview of our work in \secref{sdiscussion}. Further details and additional examples are contained in an appendix. We provide data for many of our examples online~\cite{figsharerepo} while the source of data for any remaining examples is given in our bibliography.

\section{Methods}\label{smethods}

\subsection{Basic DAG properties}\label{sdagproperties}

Suppose we have a DAG (directed acyclic graph) $\Gcal = (\Vcal, \Ecal)$ where the set of nodes and edges in the graph are $\Vcal$ and $\Ecal$ respectively. We will use $N=|\Vcal|$ for the number of nodes in the DAG. We will denote an edge from node $n$ to node $m$ as $(n,m)$. Edges can be weighted, such that a larger weight of an edge represents a stronger connection.

A \tdef{path} in any directed graph, including a DAG, is a sequence of nodes in which consecutive nodes are linked by an edge in the correct direction~\cite{N10}. The length of the path is the number of edges in that path, one less than the number of nodes in the path.
This allows us to define an \tdef{antichain} $\Acal$ as a set of nodes in which there is no path between any of the nodes in the antichain. We will later use the antichains in both the original DAG and in directed graphs we shall derive.

The lack of cycles in a DAG leads to a special property, namely that every DAG is linked to a unique \tdef{partial order} on the set of nodes.  That is we can say $n \prec m$ if there is a path \emph{from} $n$ \emph{to} $m$. Note there need not be any sense of order between two nodes, i.e.\ when there is no path between the nodes, and it is this partial order which we aim to respect in our community detection.  This mathematical property of the DAG's partial order is often seen in key properties of a real data set.
For instance, in a perfect citation network, the publication dates of documents express the partial order.  That is if  $n \prec m$ for two documents $n$ and $m$ in the citation network then we know that document $n$ must be published before document $m$ where we take the edges in the DAG to run \emph{from} the older document being cited \emph{to} the bibliographies of newer documents. The partial order in this case reflects the ``arrow of time'' inherent in the publication of documents. In computer libraries, each package tends to require less specialised packages. For a food web, the mathematical partial order may reflect the tendency of larger animals to eat only smaller ones.

In general, this sense of order in a DAG is often seen as a natural hierarchy.
For a library of computer packages, we might talk about a low-level package.
In a food web, we have ``top-level predators''. Typically, elements at the same level in the
hierarchy are going to have no connections between them since they do the same job at this level, they are alternatives to each other.  That is there is a good chance that such elements with no connections are similar and it is the order in the DAG, the level of the hierarchy is telling us about this similarity. In terms of the formal properties of the DAG, these elements form an antichain.

For example, predators of a similar size will not generally eat each other and so are often found at the same level in a food web.  Yet such animals could be similar in the sense that they compete for similar (smaller) prey. In the food web example we will consider in \ref{sfoodweb}, sharks, tarpon and grouper will be found in the same siblinarity community, because they share many common prey, such as killifish and crabs.

\subsection{Height and Depth Antichain Partitions}\label{sheightpartition}

So the approach we take is to look for communities of similar nodes in a DAG which are antichains, denoted as $\Acal$. In that way, none of the nodes in one of our antichain communities are ordered before or after any of the others in the same community.  There is no direct connection between nodes in our antichain communities by definition so these are very different communities from those produced in general networks such as discussed in~\cite{F09}.

For simplicity, we will also restrict ourselves to the case where each community is an antichain, and the set of all communities is a partition $\Afrak$, which we call an \tdef{antichain partition}.  That is every node is in exactly one element (one community) of our antichain partition $\Afrak$. This is hard or non-fuzzy clustering in the language of data science.






Our first two examples of antichain partitions are well known features of any DAG. We start by noting that another property of any DAG is that we can always assign a \tdef{height} to every node.  The height $h(n)$ of node $n$ is the length of the longest path to $n$ from any node with zero in-degree.
It is straightforward to see that each node on a path must have different heights with the height increasing as you move along the path. Conversely, nodes of the same height cannot have any path between them so that nodes of the same height form an antichain. Thus we can define the \tdef{height partition} $\Afrak^{\mathrm{(h)}}$ to be the set of antichains $\{\Acal^{\mathrm{(h)}}_a\}$, each of which contains all the nodes of a give height
\begin{equation}
 \Afrak^{\mathrm{(h)}} = \{ \Acal^{\mathrm{(h)}}_a  \}
 \, ,
 \quad
 \Acal^{\mathrm{(h)}}_a  = \{ n | n \in \Vcal, \, h(n)=a\} \, .
 \label{hacdef}
\end{equation}
Similarly, we can define the \tdef{depth} $d(n)$ of a node $n$ to be the length of the longest path from $n$ to a node with zero out-degree. Nodes with the same depth are guaranteed to form an antichain so we can define the \tdef{depth partition} $\Afrak^{\mathrm{(d)}}$ to be partition of the set of nodes by their depth,
\begin{equation}
 \Afrak^{\mathrm{(d)}} = \{ \Acal^{\mathrm{(d)}}_a  \}
 \, ,
 \quad
 \Acal^{\mathrm{(d)}}_a  = \{ n | n \in \Vcal, \, d(n)=a \} \, .
 \label{dacdef}
\end{equation}

Note that these height and depth antichains are examples of maximal antichains, each antichain is not a proper subset of any other antichain including those not in our partitions.  Many of the layering algorithms are designed to produce the same or similar numbers of antichains, so again those are often maximal antichains or something close to that. For this reason, the height and depth antichains are good representatives of the type of antichain produced by traditional layering algorithms, so they will be used to illustrate how our siblinarity antichain partitions are very different.

\subsection{Siblinarity Antichain Partitions}\label{ssiblinaritypartiton}

Often the \emph{elements} of either the height or depth antichain partitions (an element here is one community, one antichain in a partition) provide one definition of a level in the hierarchy. However, while these antichains respect the order of the DAG, we want to highlight much smaller groups which contain nodes which are much more similar than just the similarity imposed by the order as encoded by an antichain. All the nodes at one height, or those at one depth need not be very similar in general.

To add similarity to the hierarchy constraint encoded through our restriction to antichains, we can take inspiration from classic similarity measures used in bibliometrics. In that context, one way to assess the similarity of two publications is to look at overlap of their neighbours in the citation network. The more two publications share the same neighbours, the more similar they are said to be. The size of an intersection between two paper's bibliographies is called \tdef{bibliographic coupling}~\cite{K63,M64}, whereas the size of an overlap between articles that they were referenced by is called \tdef{co-citations}~\cite{S73,SG74}. So by looking into similarities of neighbourhoods of two nodes, we can say something about how similar they are themselves.

A family tree provides a good example where people are the nodes and edges are from a parent to a child and so point forward in time in terms of birth date. There will be many people in a single generation but, by definition, none will be a parent or a child of any other person in the same generation so each generation forms an antichain. Generations are layers in a natural hierarchy for this DAG. However, almost all people in one generation will have little genetic biological relationship to each other so this large antichain, a single generation, may not be very interesting in many problems.  However, if we also look for clusters of people within this generation, people who have common predecessors and so share one or two parents, then these smaller communities may be of more interest. Using such a predecessor similarity measure on top of an antichain constraint would mean a community detection method would be producing communities of siblings.


These ideas of antichain and neighbour similarity are encoded in a function which measures the quality of a given partition $\Afrak$ of our DAG into antichains, $\Acal$. Motivated by the family tree example, we call our function \tdef{siblinarity} $S(\Afrak)$ and we define it to be
\begin{equation}\label{siblinaritygen}
    S(\Afrak)
    =
    \sum_{\Acal\in \Afrak} \;\;
    \sum_{n \in \Acal} \;\;
    \sum_{m \in \Acal \setminus n}
    \left(\simfunc(n,m) - \simfuncnull(n,m) \right)
    \, .
\end{equation}
Here the first term $\simfunc(n,m)$ is some measure of the similarity of two nodes $n$ and $m$.  The second term, $\simfuncnull(n,m)$, is the expected value of similarity of these two nodes in some suitable null model. There is a lot of freedom in choosing a null model but in general it is some randomised version of the DAG.
As $m$ and $n$ are in the same antichain $\Acal$ there is no path between nodes contributing to siblinarity. We have excluded the case $m=n$ so any node in a community by itself contributes zero and $S(\Afrak)=0$ for the case where every community is a single node. Including the $m=n$ terms only adds an irrelevant overall constant.

Any similarity measure could be used but a logical choice for the similarity function in our context is the number of neighbours that $n$ and  $m$ have in common, so we will use $\simfunc(n,m) = |\Ncal(n)\cap\Ncal(m)|$ where $\Ncal(n)$ is the neighbourhood of node $n$.  Of course, we could use any other similarity measure between two sets, such as Jaccard index~\cite{J12} or cosine similarity. However, we chose to use the neighbourhood overlap, because this similarity measure is analogous to the similarity, assumed in the classical modularity~\cite{N06} and has been shown useful in studying modular structures in multiple types of networks.

There are two obvious choices for this neighbourhood: one in terms of its predecessors $\Ncalpre (n)$, as used in the family tree example above, and another in terms of the successors, $\Ncalsuc (n)$. More formally
\beq
\Ncalpre (n) = \{ m | (m,n) \in \Ecal\} \, ,
\qquad
\Ncalsuc (n) = \{ m | (n,m) \in \Ecal\} \, .
\label{Ncalpresuc}
\eeq
We could also use both, and use $\Ncal^{\mathrm{(both)}} (n) = \Ncalpre(n) \cup \Ncalsuc(n)$.

It is useful to express this in a matrix form as follows \cite{SP11}
\begin{eqnarray}\label{esiblinaritymat}
 S({\Afrak})
  &=&
   \sum_{\Acal \in \Afrak} \;\;
   \sum_{n \in \Acal} \;\;
   \sum_{m \in \Acal \setminus n}
   \big( \tilde{A}_{nm} - \frac{\kappa_n\kappa_m}{W} \big)
    \, ,
    \\
    \nonumber
    &&
     \;\;
    \kappa_n := \sum_m\tilde{A}_{nm}
    \, ,
    \;\;
    W = \sum_{n,m}\tilde{A}_{nm}
    \, .
\end{eqnarray}
Here the adjacency matrix $\vvmatr{A}$ for our DAG is defined so that $A_{nm}$ is the weight of the edge from $n$ to $m$.
The neighbourhood overlap is captured by the matrix $\Amatrtilde$ which is the product of the adjacency matrix $\Amatr$ of the DAG and its transpose.  The $\Amatrtilde$ matrix can be regarded as the adjacency matrix for a derived graph $\tilde{\Gcal}$ which is a directed weighted graph with the same node set as our DAG. In the case where we have an unweighted DAG we define this to be either $\Amatrtilde^{\mathrm{(suc)}}$, our successors-based similarity matrix, or $\Amatrtilde^{\mathrm{(pre)}}$ is a similarity matrix based on predecessors\footnote{Should we choose to use both sets of neighbours then we simply use the sum of these two matrices $\Amatrtilde^{\mathrm{(both)}}=\Amatrtilde^{\mathrm{(suc)}} +\Amatrtilde^{\mathrm{(pre)}}$ \cite{SP11}.}
where
\begin{equation}
 \Amatrtilde^{\mathrm{(suc)}}=\vvmatr{A}.\vvmatr{A}^\trans
 \, ,
 \quad
 \Amatrtilde^{\mathrm{(pre)}}=\vvmatr{A}^\trans.\vvmatr{A}
 \, .
 \label{Amattildedef}
\end{equation}
The effective similarity matrix $\Amatrtilde$ can be seen as a similarity matrix of the ``second-order'' neighbours\footnote{It is worth noting that every node is its own second neighbour so that $\Amatrtilde$ has diagonal entries (self-loops).  It is possible to exclude this effect and to replace $\Amatrtilde$ by a non-backtracking form of the adjacency matrix.  For instance we could use $\tilde{A}^{\mathrm{(NBT,suc)}}_{mn} = \tilde{A}^{\mathrm{(suc)}})_{mn} - \kout_n \delta_{mn}$ instead of $\Amatrtilde^{\mathrm{(suc)}}$. We see no strong reason to do this so we will use our simpler form.} For instance, the value of an entry $\tilde{A}^{\textrm{(suc)}}_{nm}$ is equal to the number of different walks, consisting of one step forward (with respect to the edge direction) from a node $n$, followed by one step backward (against the edge direction) such that the end node of this walk is $m$. Similarly, in $\tilde{A}^{\textrm{(pre)}}_{nm}$, the number of paths which begin with a step against the edge direction, followed by a step with respect to the edge direction, are counted between the two nodes $n,m$. By looking at this higher order structure in our DAG \cite{XWC16} we can get round the problem that there is no direct connection between nodes in our antichain communities.

The $\kappa_n$ is the strength of edges attached to a node $n$ in a graph with similarity matrix $\Amatrtilde$ and $W$ is the total strength of edges in that graph. The second term in \eqref{esiblinaritymat} also defines the null model we use.  This is a configuration model applied to the derived graph $\tilde{\Gcal}$.


The form we use in \eqref{esiblinaritymat} emulates the definition of modularity~\cite{NG04}, a similarity measure minus expected value of that similarity measure in some null model. The biggest difference between modularity and siblinarity is that we impose the antichain constraint in the communities we study.

This comparison with modularity (see~\cite{F09} for a review) suggests that we can modify sibllinarity to adjust the typical number of antichains found, the resolution of our method.  One simple method is to scale the null model term~\cite{RB06} so that
\begin{eqnarray}\label{esibresmat}
 S({\Afrak},\lambda)
  &=&
   \sum_{\Acal \in \Afrak} \;\;
   \sum_{n \in \Acal} \;\;
   \sum_{m \in \Acal \setminus n}
   \big( \tilde{A}_{nm} - \lambda \frac{\kappa_n\kappa_m}{W} \big)
    \, .
\end{eqnarray}
We will show some examples of this in the
appendix. 
However it is clear that for large $\lambda$ we expect many small antichain communities. In particular, for $\lambda \gtrsim W$ adding any node in a community by itself to any other antichain will reduce siblinarity so we expect all the antichains to be the trivial case where each antichain has just one node.  Conversely, we expect $\lambda=0$ to produce large antichain communities.

\subsection{Optimisation}\label{soptimisation}


We can use any numerical optimisation scheme to find an antichain partition that gives a value for siblinarity $S(\Afrak)$ of \eqref{esiblinaritymat} which is close to a maximum value for siblinarity. In this paper we choose to adopt the Louvain approach~\cite{BGLL08} to community detection in networks.  We chose this because it is fast and successful at finding communities  in networks and because it proved easy to adapt to our context. In this method we first use a greedy optimisation phase. In this we try moving single nodes into a different antichain community, choosing the community structure which gives the biggest siblinarity value, even if that means leaving the structure unchanged,  We sweep through all the nodes many times until there are no more changes in siblinarity.  We then initiate the second stage in which we produce a new weighted directed  network $\Hcal$ in which each antichain community in the original network is now represented by a single node in the new derived network. Edges are merged to match this new vertex set, keeping the total weight of the graph unchanged.  Note that this derived network is not necessarily a DAG, see Fig.\ B4 in the appendix for an example. However, while our discussion has been framed in terms of a DAG, the concept of an antichain is useful in any directed network with few small cycles and all our expressions, e.g.\ for siblinarity, remain valid. One then repeats the greedy optimisation with this new derived graph, first a greedy step and second a projection onto a smaller derived graph. One can stop the process with any of the derived graphs but it will terminate when there are no pairs of nodes (antichains in the original graph) which can be merged into a new antichain such that siblinarity is increased.


\subsection{Antichain Measures}\label{s:acmeasures}

Once we have found our antichains, we will need to analyse their properties and we will need some additional tools to do this for large examples.

An interesting question to ask is whether nodes in an antichain are similar to other nodes in the antichain. Many traditional methods for measuring the strength of a community are based on the number of edges between members of the community compared with edges to those outside.  Such measures fail for our antichain communities where there are no edges between community members. However, in constructing siblinarity, we have used a similarity measure based on the number of common neighbours in a suitable set of neighbours $\Ncal(n)$, such as the set of successors $\Ncalsuc(n)$ or predecessors $\Ncalpre(n)$.  Any similarity measure could be used in principle.

So for each antichain community $\Acal$ we define a similarity matrix $\Wmatr(\Acal)$ where
\beq
W_{nm}(\Acal) = |\Ncal(n)\cap\Ncal(m)| \, , \qquad n,m \in \Acal
\eeq
is equal to the number of common neighbours of nodes $m$ and $n$. There are many ways to use this. 
We can then use multiple metrics to give insights about how interconnected $\Acal$ is. For instance, we can define $W(\Acal)$ to be the average weight per node in an antichain $\Acal$,
\beq
 \frac{W(\Acal)}{|\Acal|} = \frac{1}{2|\Acal|} \sum_{n,m \in \Acal} W_{nm}
 \label{Wdef}
\eeq
which reflects on the mean similarity of nodes in the antichain. If this value is small, and the siblinarity of a given antichain,
\begin{eqnarray}\label{esibresmat2}
 S({\Acal},\lambda)
  &=&
   \sum_{n \in \Acal} \;\;
   \sum_{m \in \Acal \setminus n}
   \big( \tilde{A}_{nm} - \lambda \frac{\kappa_n\kappa_m}{W} \big)
    \,
\end{eqnarray}
is large, we can expect that the overlap of neighbourhoods of nodes in the antichain is sparse. In Section \ref{scitation} we will use these metrics as well as summary statistics to study antichains, obtained in a citation network.

\section{Data and Models}\label{datamodels}

We used a variety of models and data sets which have natural representations as a DAG in order to compare different community structures found by different methods.

\subsection{Space-Time Lattice Model}\label{stmodel}

Our first test model is a simple DAG where the nodes are placed on a square lattice at $(t,x)$ where $tL$ and $xL$  are integers between $0$ and $(L-1)$. To place edges between the node at the points of our lattice, we use the Manhattan distance
$d(n,m)=  |t_n-t_m| + |x_n-x_m|$ for the distance between nodes $n$ and $m$ of coordinates $(t_n,x_n)$ and $(t_m,x_m)$ respectively.
We add a directed edge from $n$ to $m$ with probability $p(n,m)$ where
\begin{equation}
 p(n,m) =
 \begin{cases}
  0              & \mbox{if} \; t_n \geq t_m \\
  1-\frac{d(n,m)}{D} & \mbox{if} \; t_n < t_m.
 \end{cases}
\end{equation}
The edges from $n$ to $m$ only exist if  $t_n<t_m$ and it is this arrow-of-time which ensures that we will always have a DAG.

The idea behind this DAG model is that there is a natural hierarchy given by the $t$-coordinate which guarantees acyclicity. In addition, nodes which are close in their space and time coordinates will often have large numbers of common neighbours, both successors and predecessors, whereas nodes that are distant in space or time will have few neighbours in common. So in the visualisations we expect to see nodes of the same time coordinate and close in their space coordinates, to be placed in the same antichain community.
However, as the links are placed with a stochastic mechanism, we will sometimes see nodes from neighbouring layers grouped together.  That is the antichain structure in a given realisation of our model is not a perfect match to the natural layers defined by the time coordinate.  This simulates what one finds in real data sets.  For instance, two papers written on a similar topic would be represented by nodes with a similar $x$ coordinate in this model.  If they were written independently at the same time then they could not be connected but if published at slightly different times it might be possible for the earlier paper to be cited by the later paper.

\subsection{Price Model with Subject Fields}\label{price}

One of the oldest models of citation networks is the Price model of cumulative advantage~\cite{P76} and this defines a DAG which has the fat-tailed (power-law) distribution for the number of papers with a given citation count\footnote{The undirected version of this is the Barab\'{a}si-Albert model, see~\cite{N10} for a discussion.}. Here we consider a
modified version of the Price model in which we assign nodes (representing papers) to different `fields' and we create edges (citations between papers) such that they are usually between papers in the same field. This model is too simple to capture many aspects of real citation network though it does emulate three fundamental aspects: the order of papers imposed by time, the fat-tailed citation count distribution, and the preference of most papers to cite papers within a similar field.  For our purposes, all we use here is that this gives a DAG with a well known structure and now with a planted partition. Our expectation is that our siblinarity antichain partitions will tend to cluster papers published around the same time and in the same field.

\begin{figure}
\centering
\includegraphics[width = 0.5\textwidth]{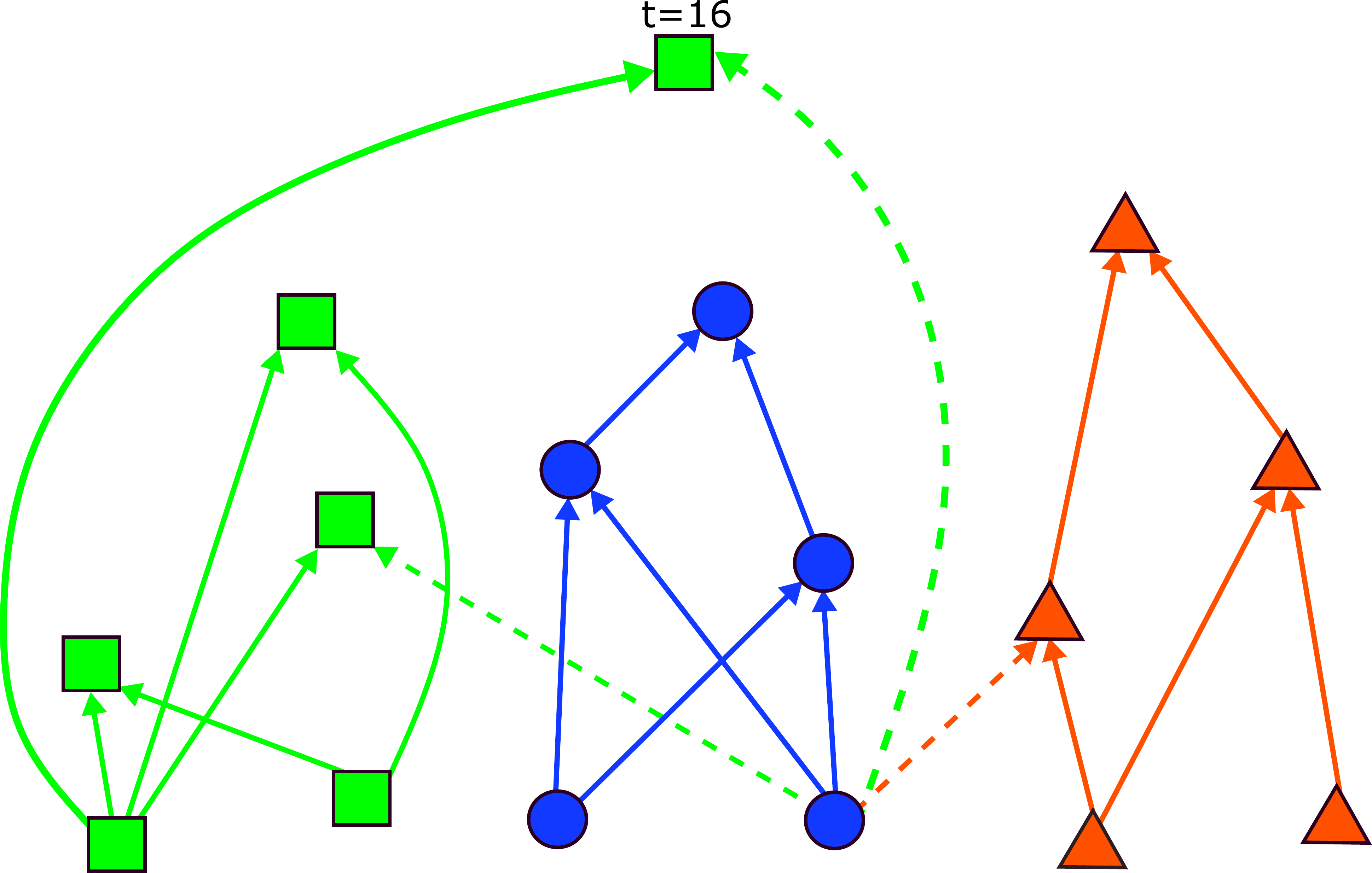}
\caption{An illustration of the Price model with two edges added per node ($m=2$) and three fields ($F=3$) as indicated by the three types of node. Time increases as you move up the diagram so the vertical ordering of the nodes represents a total order in the DAG. Solid (dashed) lines represent citations between papers of identical (different) fields. A new node labelled $16=(t+1)$ in field 0 (green squares) is added an existing network of fifteen nodes. Here we suggest the two edges added to node 16 are to the two nodes of highest degree (the cumulative advantage bias) with one in the same field (continuous edge) and one in a different field (dashed edge).}
\label{f:priceex}
\end{figure}

The model is illustrated in \figref{f:priceex}. To define this model, consider a sequence of networks $\Gcal(t)$ where $t$ is a positive integer playing the role of time and which gives us an order to the nodes in our networks. Each graph $\Gcal(t)$ has $t$ nodes. In our notation, the node $u(s)$ is always the node added at step $s$ in the process so it exists in all networks $\Gcal(t)$ provided $0<s\leq t$.
The nodes in these networks are also partitioned into different fields, that is each node $u(t)$ is in field $f(t)$, one of $F$ distinct fields.

To create the next graph in the sequence, $\Gcal(t+1)$, we first add a new node $v(t+1)$ to the vertex set. This new node also is assigned to a field, $f(t+1)$, chosen uniformly at random from the set of $F$ possible fields.

We now add $m$ directed edges to this new node $v(t+1)$ from existing nodes $u(s)$ where $s\leq t$. Following Price, we chose the source nodes $s$ for the new edges from the existing network $\Gcal(t)$ with probability $\Pi(t,s)$ defined to encode ``cumulative advantage''. That is the higher the current citation count of a paper, the more likely it is to be cited. Since we define our directed edges to run from early to late times, this means our probability $\Pi(t,s)$ is proportional to a linear function of the current out degree of existing nodes, say $\kout(t,s)$ for the citation count of node $s$ in the graph $\Gcal(t)$. For simplicity we choose Price's original form $\Pi(t,s) \propto (\kout(t,s)+1)$.

To impose a modular structure reflecting the preference of papers to cite others papers within the same field, we use a parameter $\phi$ which is the probability that a new paper $v(t+1)$ cites another paper $u(s)$ in its own field i.e.\ $f(s)=f(t)$ but where the connection is made to node $u(s)$ chosen from the papers in the same field using our cumulative advantage $\Pi(t,s)$.  With probability $(1-\phi)$, the source node $u(s)$ of a new edge is chosen from within the papers not in the same field i.e.\ $f(t+1) \neq f(s)$.

This leaves us with a stochastic model of three parameters: $m$, $\phi$ and the total number of nodes in any network we use.
We considered networks with $5,000$ nodes and $m=3$, $m=5$ edges added to each new node.
Some aspects of the model can depend on the initial graph used to start the simulation but this was unimportant for our studies.  The model is described in more detail in Appendix F.

\subsection{Real World Data Sets}\label{datasets}

To test our approach on actual data, we use three examples.

First we use the Florida Bay food-web dataset~\cite{UBE98}. In this network, the nodes are compartments and edges represent directed carbon exchange in the Florida Bay. There is an edge from $i$ to $j$ if compartment $j$ consumes carbon from compartment $i$ (often, this means species $j$ eats species $i$). The compartments are mostly organisms but also encompass special nodes such as ``input'' and ``output''. We also had group classification labels. Examples groups include ``Zooplankton Microfauna'' and ``Pelagic Fishes''~\cite{BGL16}. The original network consists of 128 nodes and 2106 edges but this contains cycles. We used the breadth first search approach, described in~\cite{SANSP17} to recover a DAG, removing 176 edges (less than 9\%). 

We also used two examples of a citation networks in which documents are nodes and we draw an edge from a newer document when it cites an older document. We used two datasets: citations between papers in the arXiv High Energy Physics-Theory repository, for papers
published between 1992 and 2003 (referred to here as \texttt{hep-th})~\cite{KDDcup} and a selection of Cora~\cite{LG03,SNBGGE08}
papers, referred below as \texttt{cora}. The \texttt{hep-th} dataset contains a set of
27,770 papers and 351,500 edges, that represent citations between the papers. The \texttt{cora} dataset consists of a selection of 2,708 scientific publications on Machine Learning, classified into one of seven classes, based on their topics. The papers were selected in a way such that in the final network every paper cites or is cited by at least one other paper. This citation network consists of 5,429 links\footnote{In practice there are often ``bad'' links which are in the ``wrong'' direction, from a newer document to an older document. This is because documents are published in different versions and the text available may not have been created at the time associated with the document in the data
set. For instance, a revised version of an arXiv paper carry the same index as the first version. A journal article has several associated dates:
first submitted, date accepted, published online, formal publication date and so forth. Such bad links can introduce cycles and these must be
dealt with. For our arXiv data we simply dropped the links that do not respect the agreed order of time; they account for less than 1\% of the data. For the \texttt{cora} data we used the approach, discussed in~\cite{SANSP17} to recover a DAG from a directed network, leaving us with a network, composed of 5,255 edges.}.
We use a convention that a citation of a paper $u$ by a paper $v$ is represented by an edge ($u,v$).

\section{Results}\label{sresults}

\subsection{Space-Time Lattice}

\begin{figure}[!htb]
\centering
\includegraphics[width = 1\linewidth]{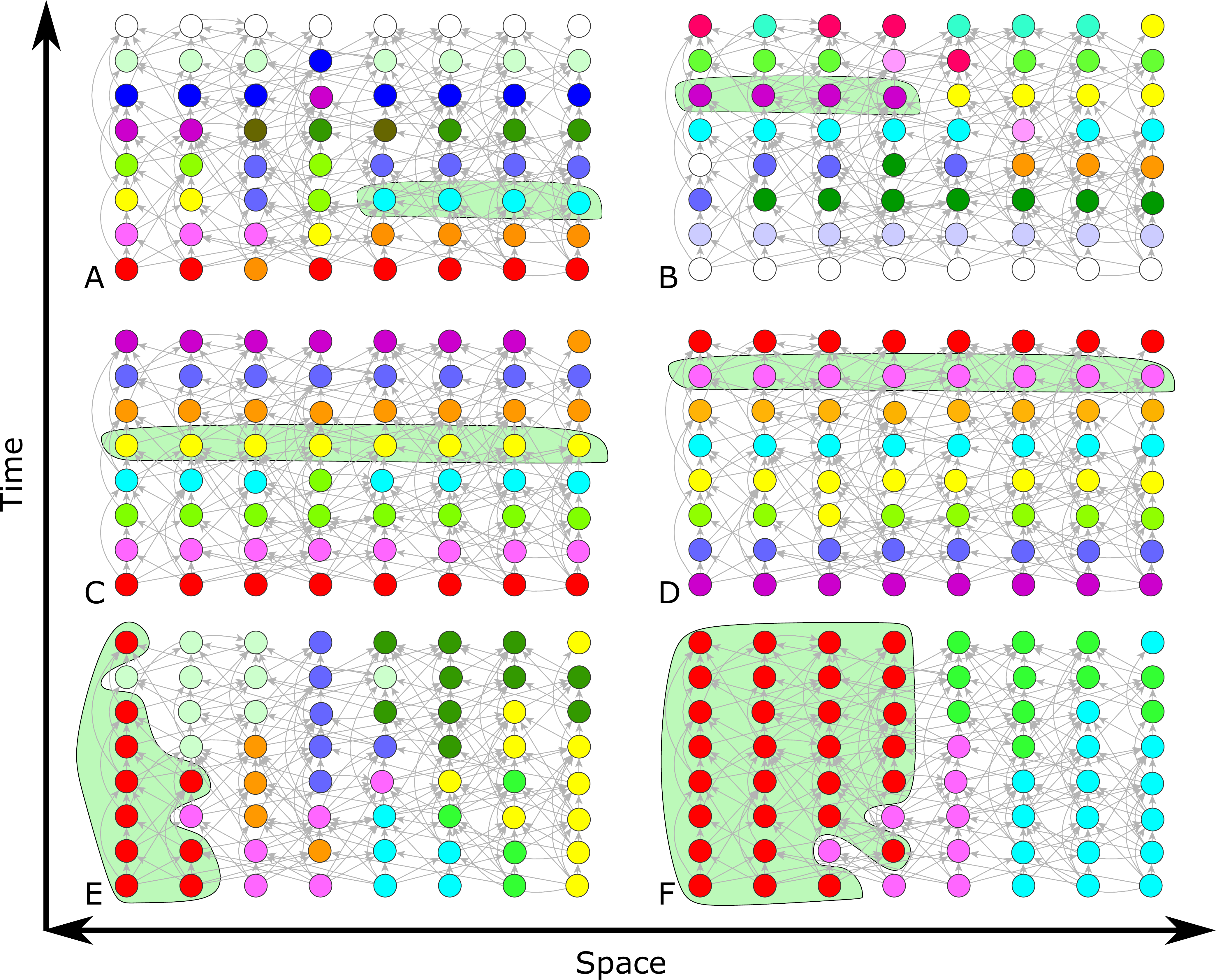}
\caption{Examples of various partitions of DAG from a simple space-time model. In this DAG, edges are more probable between two nodes if they are a shorter Manhattan distance apart (in the space-time). The time coordinate, which induces the acyclicity in the network, is vertical. Colour indicates community of a node. The top row shows siblinarity communities based on common successors (A), and predecessors (B). The central row shows layers based on height (C) and depth (D). The communities in the bottom row are from modularity community detection using resolution parameter value of one (E) and two (F). Nodes coloured white are in a community of size one. Siblinarity partitioning tends not to form communities (top row) which stretch across the whole network unlike those communities based on height or depth shown in the centre row. The communities found using traditional methods respect the spatial (horizontal) constraint but show no respect for the order in the DAG as they are spread over several times vertically. These key differences are exemplified in the figure by highlighting one representative community in each partition (green cells).}
\label{fstdag}
\end{figure}

We use the space-time lattice model of \secref{stmodel} on a small scale because we can make effective visualisations which illustrate the key ideas of communities in DAGs. In \figref{fstdag} we show the same instance of the space-time lattice model in which nodes are placed on an eight-by-eight square lattice. We then use colour to show the different partitions produced using different algorithms.

The bottom two examples of  \figref{fstdag} illustrate standard community detection algorithms applied to our network. Here these are found by optimising modularity with two different resolutions (obtained by scaling the null model term). The communities found tend to be located in one region of space reflecting the spatial constraint in the model. However, they extend over several values of time so these communities do not respect the order inherent in the DAG.  That is, the nodes in the these modularity communities are not similar in terms of their time coordinates.

If this was a citation network, we would be grouping papers together with different publication dates.  If we were interested in comparing the impact of each paper through citation counts, the older papers would have an advantage making the comparison unfair.
If this toy model represented a food web, these communities derived from the undirected network might capture aspects such as distinct parts of the ecosystem.  However it fails to find the hierarchy, each group would contain predators and their prey.

So these communities from the undirected graph may be of use in some contexts, but they are not particularly sensitive to the inherent structure of the DAG coming from the time direction.

The middle two examples in  \figref{fstdag} use the height- and depth-antichain partitions.  Now the hierarchical structure coming from the time direction is clearly exposed, with many communities running across one row, perhaps two. Where two rows are involved, it is showing how the these community detection methods are highlighting where the placement of the nodes in the visualisation does not reflect the topological reality because of the stochastic aspect of edge placement.  This shows that these antichain partitions can do a useful job, picking out the true topological hierarchy as defined by the data.

However the height- and depth-antichain communities show the opposite problem from the undirected graph communities.  That is they respect the order coming from the arrow of time but they fail to pick up in any way the spatial clustering in the data.
If this was a citation network, these would cluster papers published at a similar time regardless of academic field.  
For a food web, all predators at the same level would be grouped together regardless of whether or not they were competing in the same ecological niche.

The top two networks  in  \figref{fstdag} show the siblinarity antichain communities, one based on predecessor neighbours and the other using successors.  This shows that the communities respect both the time and the spatial clustering in the data. The nodes tend to be in the same row and only contain nodes which are close by reflecting the spatial clustering in the model. This suggests that when applied to a citation network, this method will cluster papers which are directly comparable, similar publication date and similar field.
Applied to food webs, siblinarity will only group predators at the same level competing in the same niche.

\subsection{Diversity of Antichains: The Price Model with Fields}\label{sdivpricemodel}

The space-time model illustrated that the communities formed by our methods respect the order implicit in a DAG. Our method also aims to create groups which are similar in other ways as reflected in the network structure. To test this we use our modified version of the Price model of \secref{price}. We will use the language of a citation model, reflecting Price's original context. So here we say we are aiming to group papers (nodes) of a similar age (in an antichain) published in the same academic field based on the network structure alone (using co-citation or bibliometric coupling).

Here we consider the case when a field is assigned stochastically using a uniform distribution so that it is equally likely to assign a field $f=1$ and $f=5$.
Other variations are possible, for instance, one could assign the fields sequentially deterministically or create non-uniformity in field sizes. Some of these variations will be discussed in Appendix F.

Our aim is to show that our antichain partitions are largely composed of papers from the same field which we measure using Shannon's diversity metric $D$~\cite{J06}.
That is given an antichain $\mathcal{A}$ we find $p_f$, the fraction of nodes in the antichain which are in field $f$.  The diversity of this antichain is then given by
\begin{equation}
 D(\Acal)
 =
 \exp \left( - \sum_f p_f \ln(p_f)   \right)
 \label{e_diversity}
\end{equation}
so that $1\leq D(\Acal) \leq |\Fcal|$.

\begin{figure}[h!]
\centering
\includegraphics[width = 0.9\linewidth]{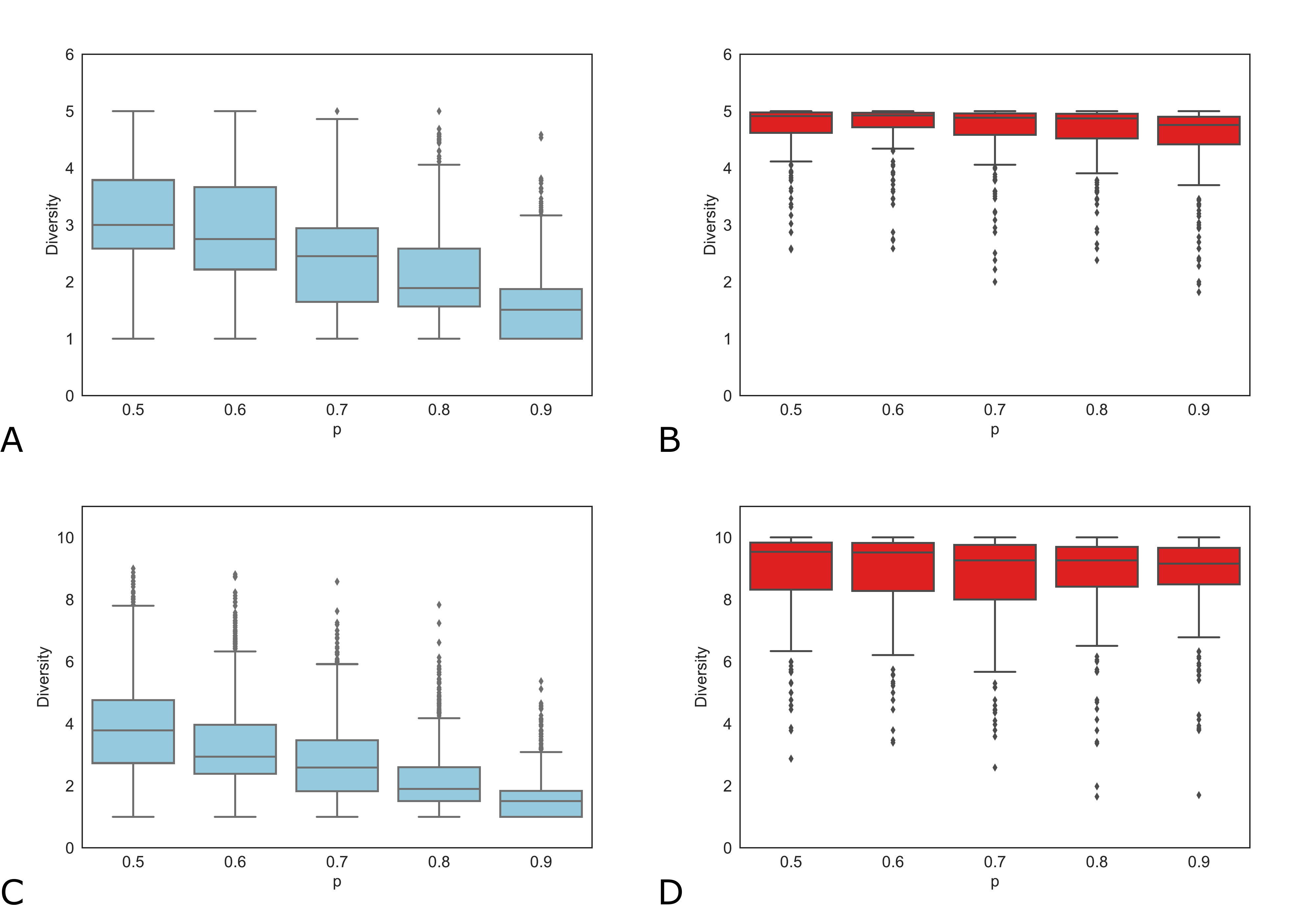}
\caption{Comparison of diversity in height-based antichain partitions (red boxes on right) versus siblinarity communities, based on common successors (blue boxes on left) in the Price model with five fields (top row, Fig.\ A and B) and ten fields (bottom row, Fig.\ C and D) and different intraconnectivity probability $p$. We studied ten networks of  $5,000$ nodes for each set of parameters. Figures show that the diversity of siblinarity antichains is clearly significantly smaller than that of heights. The box extends from the lower to upper quartile values of the data, with a line at the median. The whiskers extend from the box to show the range of the data. Flier points are those past the end of the whiskers.}
\label{faveragediv}
\end{figure}

\begin{figure}[!htb]
\centering
\includegraphics[width = 0.7\linewidth]{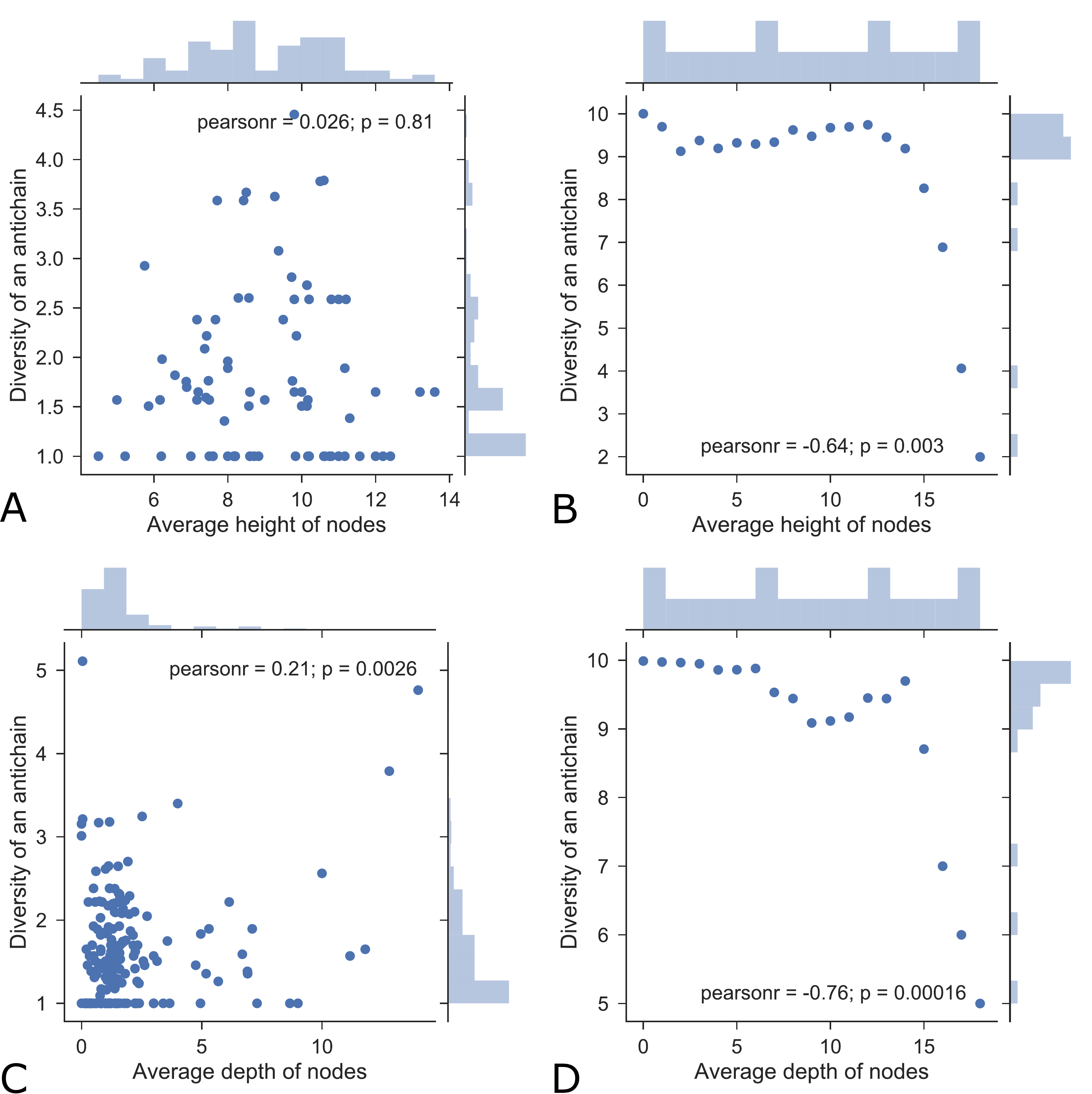}
\caption{Shannon's diversity of antichains in a Price model with planted partition of nodes into three fields for networks of $5,000$ nodes. Each node  is assigned to a field stochastically using a uniform distribution and nine times out of ten a new node chooses to connect to an existing node of the same field, with a preference for large degree nodes, $\phi=0.9$. Each new node attaches to $m=3$ older nodes. Each point represents an antichain in a given partition, with height or depth and Diversity $D(\Acal)$ \eqref{e_diversity} used as coordinates. Four different types of antichain partition are shown: A - successors-based siblinarity antichains, B - height-based antichains, C - predecessors-based siblinarity antichains, D - depth-based antichains. The height and depth antichains are clearly much bigger (few points on their plots) but much more diverse, with their diversity close to the theoretical maximum $10.0$. On the other hand, the siblinarity antichains based on co-citation similarity  (A) or bibliometric coupling (B) are drawn largely from the same field with diversities close to the minimum value of $1.0$.}
\label{f_pricewfields_shannon}
\end{figure}

The results for the average diversity of a partition, $\sum_{\Acal\in\Afrak}D(\Acal)/|\Afrak|$ are shown in \figref{faveragediv}, results for the distribution of $D(\Acal)$ values within $\Afrak$ are shown in \figref{f_pricewfields_shannon}. Not surprisingly the diversity of height and depth antichains is almost always larger than that of communities found from a siblinarity antichain partition. This is expected because the height and depth antichains typically contain more nodes, reflected by the low number of points on their plots, as their construction is not affected by the field labels. Since they are constructed without regard to the field labels, we would expect, and find, that the diversity of the height and depth partitions is close to the maximum value where $p_f \approx 1/|\Fcal|$ and so $D \approx |\Fcal|$.

We also noted that the diversity of antichains that are height or depth based does not approach maximum if the sizes of fields are not uniform. This is expected as Shannon's entropy not only accounts for number of species in the ecosystem but also their relative abundances. Uneven sizes of fields result in expected diversity score lower than maximum, as the maximum would be obtained if we had equally abundant species, as seen in the other two figures. 

\clearpage


\subsection{Florida Bay food web}\label{sfoodweb}

We also studied our antichain partitions in the DAG version of the Florida Bay food web data set~\cite{UBE98}.
Results are shown in \tabref{tfoodwebboth}.

\begin{table}[htb!]
\footnotesize
\centering
\begin{tabular}{c|p{0.9\textwidth}}
No.\ & Species (type)\\
\hline\hline
1 & $2\mu m$ Spherical Phytoplankt, Synedococcus, Oscillatoria, Small Diatoms ($<20\mu m$), Big Diatoms($>20\mu m$), Dinoflagellates, Other Phytoplankton, Benthic Phytoplankton, Thalassia, Halodule, Syringodium, Roots, Drift Algae, Epiphytes.\\
3 & Acartia Tonsa, Oithona nana, Paracalanus, Other Copepoda, Other Zooplankton, Sponges, Bivalves.\\
7& Coral, Other Cnidaridae.\\
9& Benthic Crustaceans, Detritivorous Amphipods, Herbivorous Amphipods, Detritivorous Gastropods, Detritivorous Polychaetes, Suspension Feeding Polych, Macrobenthos, Detritivorous Crabs.\\
18& Toadfish, Brotalus.\\
19 & Other Killifish, Goldspotted Killifish, Blennies, Clown Goby, Silverside, Lobster, Predatory Crabs, Callinectus sapidus, Bay Anchovy, Rainwater killifish, Mullet, Other Horsefish, Gulf Pipefish, Dwarf Seahorse, Code Goby, Halfbeaks.\\
22 & Flatfish, Grunt, Pinfish, Rays, Porgy, Scianids, Parrotfish, Bonefish, Needlefish, Snook, Puffer, Manatee.\\
23 &Omnivorous Crabs, Pink Shrimp.\\
25 & DOC, Isopods, Herbivorous Shrimp, Thor Floridanus, Sailfin Molly,
Green Turtle.\\
26 &Sharks, Tarpon, Lizardfish, Grouper, Jacks, Pompano, Gray Snapper, Red Drum, Mackerel, Small Herons \& Egrets, Ibis, Roseate Spoonbill, Herbivorous Ducks, Omnivorous Ducks, Gruiformes, Small Shorebirds, Gulls \& Terns, Kingfisher, Loggerhead Turtle, Hawksbill Turtle.\\
27 & Other Snapper, Other Pelagic Fishes, Spotted Seatrout.\\
28 & Stone Crab, Sardines, Anchovy, Other Demersal Fishes, Filefishes.\\
29& Barracuda, Loon, Greeb, Pelican, Comorant, Big Herons \& Egrets, Predatory Ducks.\\
33 & Free Bacteria, Dolphin.\\
35 & Raptors, Crocodiles.\\
36 & Output, Respiration.\\
Ind.&  Input (0), Water Flagellates (2), Water Cilitaes (4), Benthic Ciliates (5), Meroplankton (6), Meiofauna (8), Benthic Flagellates (10), Water POC (12), Predatory Gastropods (13), Echinoderma (14), Predatory Shrimp (15), Predatory Polychaetes (16), Mojarra (20), Benthic POC (24), Spadefish (31), Catfish (32), Eels (34).\\

\end{tabular}
\caption{Siblinarity communities, based on common prey and common predators in the Florida Bay food web. Some communities consist of a single species and these are listed together under  community number ``Ind.'' to reduce the size of the table. The numbers in brackets indicate the community index in our dataset. Note the green turtle in community 25 appears to be out of place with the other smaller species in the same community but they all tend to feed on similar species.}
\label{tfoodwebboth}
\end{table}


We see that many of our siblinarity antichains consist of similar species as indicated by their similar names. For instance community 9 consists of, amongst other crustaceans, several types of amphipods. Another example of how our siblinarity antichains work is the example of the green turtle in community 25. The ``Green Turtle" node has been placed in a community along with much smaller species, seemingly very different from turtles, such as shrimp and isopods. However, green turtles feed on thalassia (a type of seagrass, commonly known as turtlegrass), which is also a food source for organisms, represented with ``DOC'' (Dissolved Organic Carbon) node and isopods. Furthermore, all of the species in this green turtle antichain community feed on epiphytes. Nodes in this antichain all have the same height of two, but their depths range from 22 for ``Epiphytic Gastropods''
 to 30 for ``Isopods'', ``Herbivorous Shrimp'', and ``Thor Floridanus''. Another example of an antichain, in which nodes of a variety of heights and depths were collected into a siblinarity community is the community 27, consisting of ``Other Snapper'' (height 27, depth 4), ``Other Pelagic Fishes'' (height 28, depth 4) and ``Spotted Seatrout'' (height 29, depth 3).

\subsection{Citation networks}\label{scitation}

The Science Citation Index, introduced by Garfield in 1964, has a functionality to search for ``Related Records''~\cite{N10}. A Related Record is any record which shares at least one cited reference with the original source record. The more shared references, the more closely related the records are --- an extension of the notion of citation searching to track a subject area. The related records are further ranked based on the number of shared references.

So in a bibliometric context our approach is an extension for ``Related Records'' functionality in the SCI. As pointed out by Newman, simply using bibliographic coupling or co-citations is flawed: strong bibliographic coupling or co-citations only occur between papers that have either large bibliographies or are highly cited, respectively~\cite{N10}. Our approach naturally eliminates this flaw: by comparing the observed neighbourhood overlap to a null model, we can evaluate the significance of the similarity regardless of the degrees of two nodes.

To understand what type of nodes are joined together in antichains and why, we looked at various statistics for successor siblinarity antichains in the \texttt{hep-th} dataset. These neighbours are newer papers that cite the papers in the antichain communities. We found 10,806 successor based antichain communities, 7,352 of which are composed of singular nodes and 1,225 are composed of at least five nodes. Several of the measures we found useful can be defined in terms of a bipartite network constructed for each antichain community.  This is simply the undirected subgraph of the full network, containing the nodes in the antichain of interest, the nodes of all their neighbours (here successor neighbours) and then all the edges between these nodes in the original graph. This is a bipartite network given the two types of nodes (see Appendix D for a formal definition). To remove noise from weakly connected nodes in antichain communities, we also found it was useful to reduce the bipartite subgraphs by dropping any neighbour node that is connected to just one node in the antichain.

We then applied a number of simple network statistics to these bipartite subgraphs in order to understand the nature of our antichain communities.  Our results are summarised in \figref{fhepth_succ_antichain_stats}. Several simple network statistics, formally defined in Appendix D, proved useful. The most basic measures are the size of each antichain size $|\Acal|$, and $|\Ncal|$ the total number of neighbours of each community. The ratio of these gives us the average number of neighbours of an antichain node in each community, $\langle k\rangle$, and we also use the standard deviation of that measure, $\sigma(k)$. A low $\langle k\rangle$ indicates a sparse ``zig-zag'' pattern with very weak overlaps between members of the antichain. A large $\sigma(k)$ indicates that we may have one well connected node in the antichain plus many connected to only a couple of neighbours. The last network statistic shown in \figref{fhepth_succ_antichain_stats} is the density of the bipartite graph, the total number of edges divided by the maximum number possible which is equal to $\langle k\rangle/ |\Ncal|$. This final network measure is close to one only if almost every member of our antichain community has the same set of neighbours.

The majority of the successor antichain communities we find in \texttt{hep-th} are small, composed of around $10$ to $20$ nodes.  They typically share around $40$ neighbours though a few antichain communities can have close to a 1,000 shared neighbours. The distribution of $\langle k\rangle$ peaks around 6 but there is often a sizeable variation in the degree as shown by similarly large values of $\sigma(k)$.

The density of the bipartite graphs, the ratio between  $\langle k\rangle$ and $|\Ncal|$ in
\figref{fhepth_succ_antichain_stats},
shows that the antichain communities in \texttt{hep-th} possess a wide range of values. The distribution of the density peaks at values close to $0.25-0.5$ for many antichains, but the number of antichains with minimal and maximal values of the ratio is non-negligible.


Our last statistic is not a network measure but exploits information we have on the month of publication for each paper in our \texttt{hep-th} dataset.
We looked at the standard deviation in the ages of papers in each antichain. We see that nodes that feature in the same antichain are also nodes of similar age, which varies by less than one year for most antichains.

\begin{figure}
    \centering
    \includegraphics[width=\linewidth]{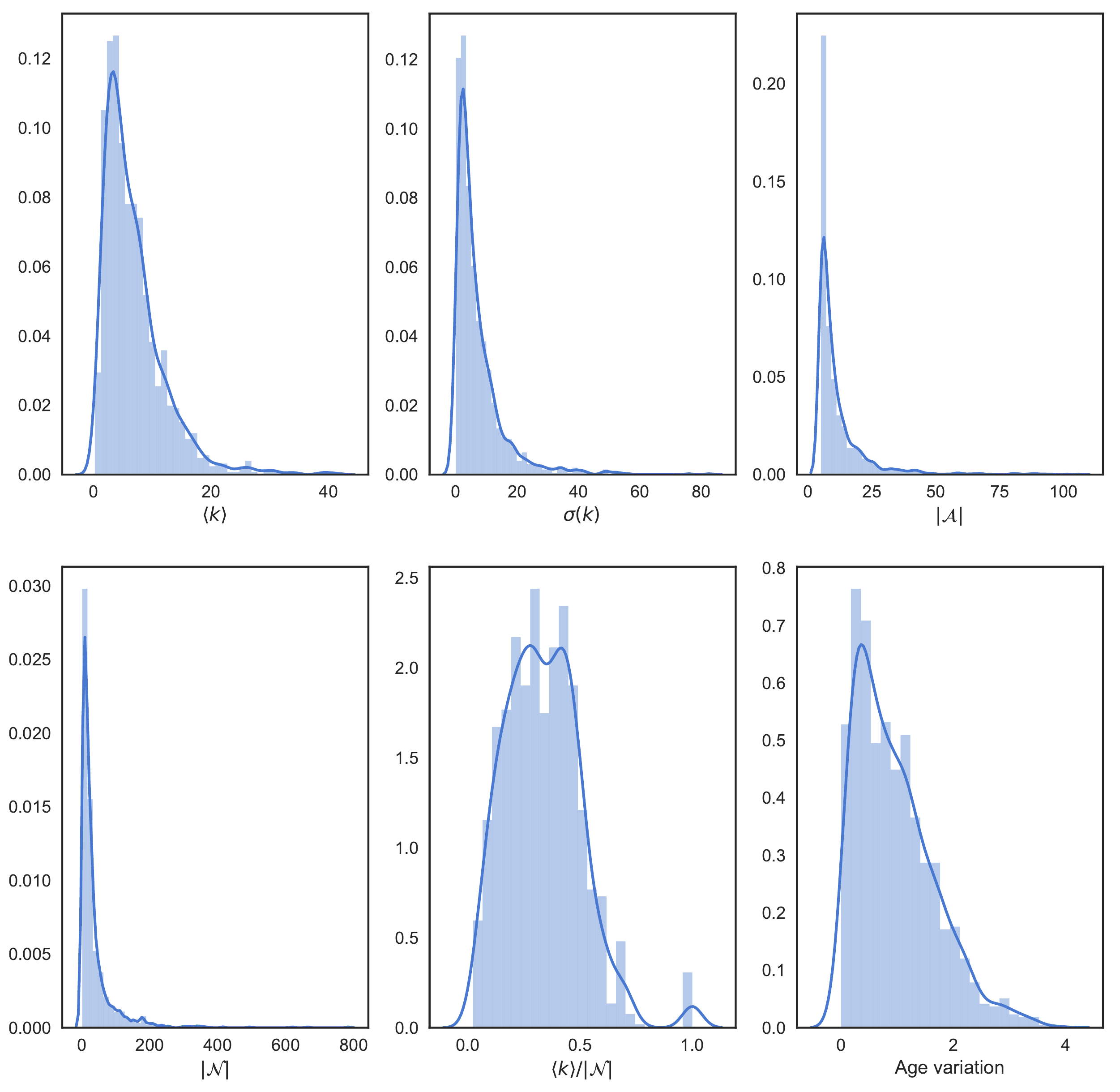}
    \caption{Summary statistics of antichains and shared successors of those antichains, based on which they were obtained in \texttt{hep-th} citation network. In each case the relative frequency distribution is plotted over all the antichain communities found using successor neighbours, younger neighbours who cite the older papers in the antichain community. Note that only those antichains, that are composed of at least five nodes are included in the statistics.
    The statistics used are as follows:
    the average number of neighbours of each antichain node $\langle k\rangle$, the standard deviation of the number of neighbours of each antichain node $\sigma(k)$,
    the size of each antichain community $|\Acal|$,
    the density of each bipartite graph $\langle k\rangle/ |\Ncal|$. The final panel shows the age variation of papers in each antichain.  That is the standard deviation in the publication date of papers in each community, known in terms of months but we give the plot in terms of years.}
    \label{fhepth_succ_antichain_stats}
\end{figure}

We also looked at siblinarity values obtained by antichains in \texttt{hep-th} and how they relate to $W(\Acal)/|\Acal|$, the mean similarity of nodes in an antichain as defined in \ref{s:acmeasures}, as well as the size of antichains $|\Acal|$.

\begin{figure}
    \centering
    \includegraphics[width=0.7\linewidth]{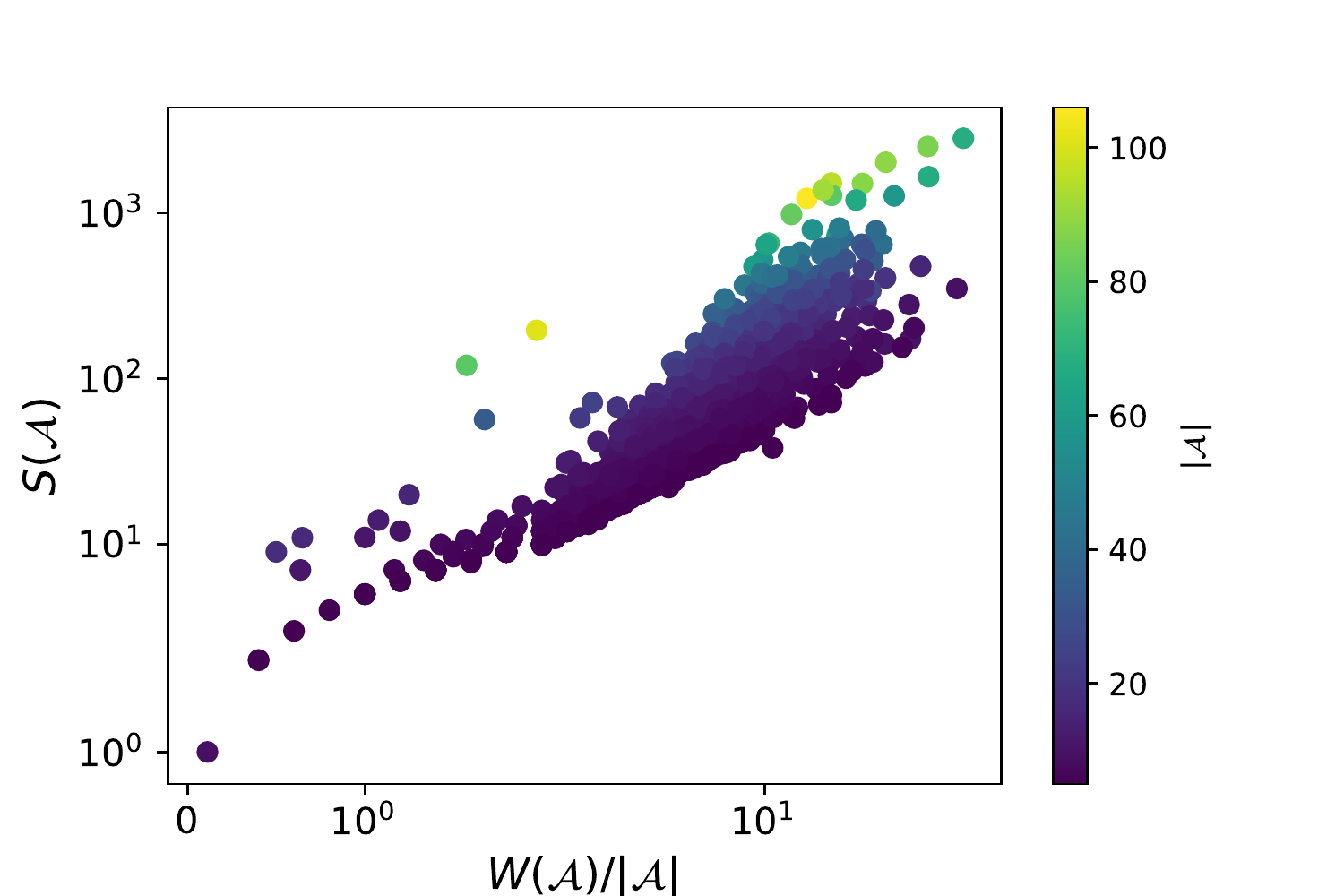}
    \caption{A colour plot showing a relation between siblinarity score of an antichain $\Acal$, $S(\Acal)$, $W(\Acal)/|\Acal|$ and $|\Acal|$ for an antichain partition obtained in \texttt{hep-th} citation network. Antichains were obtained considering co-citations. Figure shows that siblinarity scores, obtained by antichains relate to $W(\Acal)/|\Acal|$ but not to the size of an antichain: small antichains obtain siblinarity scores, larger than those of much bigger antichains. }
    \label{fhepthsiblinarity}
\end{figure}

\Figref{fhepthsiblinarity} shows that for a given size of an antichain community, there are a wide range of values for the average neighbour overlap $W(\Acal)/|\Acal|$ and the siblinarity measure. Higher overlap tends to produce a better quality antichain, that is a higher siblinarity value, but there is still a large variation, reflecting the role of the null model.

For instance some large antichains have small siblinarity scores. Conversely, the largest siblinarity score is obtained from an antichain community which is composed of 68 nodes (antichain with an index 2175 in our data), large but not the largest we found. This antichain is composed of papers, published in 1992--1994, some of which are relatively highly cited (large out-degree). The paper with the largest out-degree of those in this community has an out-degree of 112 (paper index $9201061$).

\begin{figure}[!h]
    \centering
    \includegraphics[width=1\linewidth]{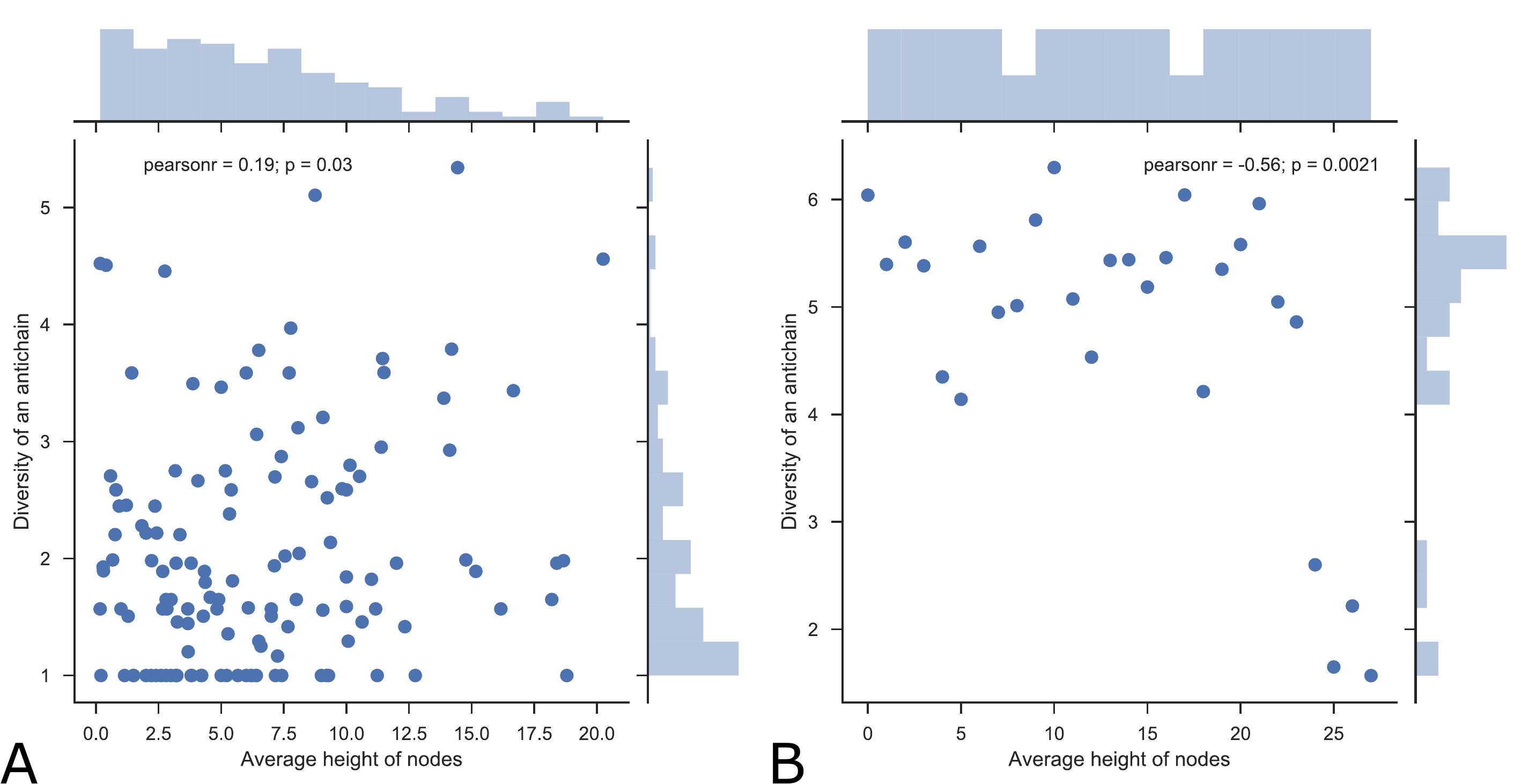}
    \caption{Shannon's diversity of nodes in antichains of the \texttt{cora} citation network. Each node has a label assigned to it, based on the topic the paper is on. There are seven topics in total. 
    Two different antichain partitions are shown: A --- successors and predecessors-based siblinarity antichains, B --- height-based antichains (diversity calculated only for antichains, that are composed of at least five nodes). The horizontal axis is average heights of nodes in the antichain. The height antichains, with diversity values often close to the maximum value of seven, are much more diverse than the siblinarity communities.
    }
    \label{fcoradiversity}
\end{figure}

We also studied a second citation network, the \texttt{cora} dataset, in which each node is labelled by one of seven different topics.  We use this label to quantify the diversity of the nodes in the antichain communities we find, similar to the analysis in Section \ref{sdivpricemodel}. \Figref{fcoradiversity} shows that the diversity of 134 siblinarity antichains is much smaller than the partitions based on heights. Furthermore, most of the time the diversity values for siblinarity antichains are equal to one, which indicates that all nodes in the antichain have the same label. This result confirms that the siblinarity communities are communities of similar papers, not just ones published around the same time.

\clearpage
\section{Discussion}\label{sdiscussion}


In network analysis an edge between nodes generally indicates a close relation and a similarity between the two connected nodes. This is exploited to understand local features, such as centrality of individual nodes, as well as gain insights on a macroscopic structure, such as the extent of community structure.

In this work we have highlighted that in some circumstances, the absence of a pairwise relationship can be a just as useful a signal as the presence of it. Our focus has been on Directed Acyclic Graphs (DAGs) where the implicit order in such networks prohibits the direct connection of many similar nodes. As this order is intrinsic to the very nature of a DAG \cite{KN09,CE14,C14a,CE16,CGLE14,ECV19}, our response has been to embrace this order as it reflects important features of the data. To find meaningful communities, we have turned to antichains, sets of disconnected nodes which often play a useful role in the study of DAGs and which reflect the topological properties on the networks we consider.

However, communities in networks can be defined as nodes which share strong intra-community connections \cite{F09}.  Since the strongest connection is a direct edge between members of nodes in the same community, the antichain seems to be the precise opposite of a network community. However, even in tightly connected network communities, not all nodes are connected to each other.  So network communities reflect on a larger mesoscopic scale of intra-connection in a network. So to find antichain communities we look beyond one edge and use node-node relationships as defined by pairs of edges in order to find closely related nodes. That is to ensure our antichains contain similar nodes, we have used neighbourhood overlap to define the siblinarity of a partition of a DAG into antichain communities. Neighbourhood overlap has long been used to define node similarity based solely on the network topology and it does not require a direct connection between nodes.

We have shown using many examples how the antichains which maximise siblinarity are interesting and relevant.
Our communities are extremely different from usual communities found in network analysis as the simple examples in \figref{fstdag} show. The communities in traditional network analysis consist of strongly interconnected nodes while siblinarity communities consist of nodes that are strictly not connected, but which share similar neighbours.

Likewise, it is important to note that our siblinarity antichains are very different from those produced by existing antichain methods such as the graph layering method often used in visualisation \cite{STT81,M71,TH13,GKNV93,NT06,HN02,HN13}.  The height or depth antichains used in our work represent such layering methods and are used to illustrate the difference between those methods and our approach. These graph layering methods methods have different goals which, directly or indirectly, produce as few antichains as possible, often ``maximal antichains'' which are not proper subsets of any other antichain.  As a result such layering algorithms produce communities in which the nodes have very different properties, the antithesis of the goal in community detection \cite{F09} which is to find clusters of similar nodes.

However, as with most community detection methods, we have a resolution parameter, our $\lambda$ in \eqref{esibresmat}, which can change the size of typical communities.  In our case, lowering the value of $\lambda$ decreases the number of antichains found so we can interpolate from antichains of very similar nodes of primary interest to us, to a DAG broken into a small number of layers regardless of the similarity of nodes within those layers. In that limit we expect to produce something similar to traditional layering methods.

The resolution parameter highlights another issue. Our main aim was to illustrate how we could reconcile the order constraint implicit DAG with the strong intra-community connections typical of network clustering methods. However, we have illustrated our concept using a modularity-like construction, but modularity brings its own baggage such as a well known resolution problem \cite{FB06}. Our resolution parameter, or similar resolution parameters \cite{SDYB12,LDB14}, can be exploited to find ``better'' communities as defined by suitable measures, but we are also mindful that there is some debate about how to define a ``good'' community structure e.g.\ \cite{PLC17}. To see if the ideas and solutions that worked for undirected graphs transfer well to the antichain communities we have suggested for DAGs, it is probably best to work within one area.  Citation networks often come with rich metadata necessary for such studies, perhaps complementing traditional approaches such as those discussed in \cite{GGS17}.

This novel type of node partitioning has several potential uses whenever data is well represented by a DAG.
In the context of bibliometrics, it is extremely difficult to measure the impact of papers published at different times in different fields. Different research fields publish papers at different rates and cite at very different frequencies, for instance pure maths and astrophysics have very different citation patterns.  Assigning a field to a set of papers is non-trivial, for example see \cite{BGGHSTEVW17}.
Even the publication date is not unique \cite{HBC15} as an article has a range of dates: the appearance of various electronic versions, the date it was assigned an index number, formal publication date, and so forth.
What our approach offers is the ability to group truly similar papers together, in terms of both topic and age, using the topology of the citation network alone.  For instance it is sensible to compare the impact of papers in our antichains as the effects of time and field have been accounted for and the papers should have a similar opportunity to make an impact. Other approaches tend to rely on metadata such as ``publication dates'' (unreliable as noted) so our approach using the citation network alone complements and enhances the tools already available. Another use could be as a recommender system for publication search: given an input paper, one could find alternative similar publications by looking into the antichain of the input paper.



Our siblinarity algorithm, like any clustering algorithm, can also be used to coarse-grain a DAG which can help navigate a large amount of data. In our case we would contend the order of a DAG is important.  Having a predator and its prey in the same community may not be a useful classification when examining food webs. Indeed we exploit this coarse graining in our implementation of the Louvain algorithm to find antichain partitions which have approximate maximal values of siblinarity.

More broadly, we see our work as emphasising that the order found in some data is an important feature which should not be ignored.  However, in the case of DAGs, this constraint is not reflected in most traditional network measures suggesting that network analysis methods need to be reexamined carefully in such cases. This has been acknowledged in some contexts, the use of co-citation and bibliometric coupling, our neighbour overlap similarity measure, is well known in scientometric analysis and again it is a response to the idea that many similar papers are not always directly linked by a citation. However, enforcing the constraint has not always been taken to its logical conclusion as it requires further adaptation of existing methods.


We have developed our work in terms of a directed acyclic graph.  However, the natural network representation of many data sets where there is a strong order or hierarchy is a directed graph with a few cycles. For instance in citation data, a document can cite a document with a later publication date. In part this is because documents have several different publication dates~\cite{HBC15}. Such backwards citation links are regularly found, with between 0.005\% and 0.4\% of links reported to be in the wrong direction in various citation networks~\cite{CGLE14}.

There are, however, several methods to produce an exact DAG from such data which then allows our methods to be applied. One could just delete the few links which do not respect the dominant order~\cite{CGLE14} or one could use more sophisticated approaches such as ``agony''~\cite{GSLMI11,T17,LBL18}. Our antichain method suggests another approach.  Our method works in principle on directed graphs, not just those with no cycles.  We exploit this in the coarse graining step of our Louvain-style numerical optimsation, which can produce cycles in the derived graphs used at later stages of the method, see Appendix C. Our approach should work well if there are few cycles.  In this case, one can look at ``bad links'', cases where citations go from an older to a newer document or perhaps even go in both directions. The two documents connected by such a bad link will be placed in two different antichain communities.  By looking at the properties of the two communities, we might be able to find the best direction for the bad link. We could even see if deleting the bad link makes most sense as indicated by a higher siblinarity score for the merger of the two antichain communities.

It is worth pointing out that although we developed our methodology in terms of DAGs, this type of partitioning can be applied to other types of networks, such as multilayer networks, bipartite networks, directed networks and even undirected networks (although in the case of the last type, antichains would most likely be composed of individual nodes). Higher order structures, such as our two-step network used in our version of modularity, are important in many areas of research so this could be a way to tackle more general network problems such as those highlighted in \cite{TWE19}.




\section*{List of abbreviations}

 DAG --- Directed Acyclic Graph

\section*{Declarations}

\subsection*{Availability of data and materials}

Our code implementing these methods, the network data analysed during this study, and the datasets supporting the results discussed in this article are all available in a Figshare repository \tsedoi{10.6084/m9.figshare.9725159}~\cite{figsharerepo}.

\subsection*{Competing interests}
  The authors declare that they have no competing interests.

\subsection*{Funding}

This work was funded by EPSRC, grant EP-R512540-1.

\subsection*{Author's contributions}

Both authors developed the theoretical concepts and designed the experiments discussed in the paper. V.V.\ created the software used and performed the data analysis.
Both authors wrote the manuscript.
T.S.E.\ supervised the work.

\subsection*{Acknowledgements}
  V.V.\ acknowledges financial support from EPSRC, grant EP-R512540-1.

\newpage


\appendix
\renewcommand{\thesection}{\Alph{section}}
\renewcommand{\theequation}{\thesection\arabic{equation}}
\renewcommand{\thefigure}{\thesection\arabic{figure}}
\renewcommand{\thetable}{\thesection\arabic{table}}
\setcounter{section}{0}
\setcounter{equation}{0}
\setcounter{figure}{0}
\setcounter{table}{0}

\begin{center}
{\Large\bf Appendix}
\end{center}

\section{Siblinarity antichain partition}\label{smethodssiblinarity}

We require a function which measures the quality of our partition of the set of nodes into our ``siblinarity antichains''.  There are two main aspects to such a function: imposing the antichain constraint and using defining node similarity using neighbourhood overlap.

Consider a directed graph (digraph) $\Gcal = (\Vcal, \Ecal)$ where $\Vcal$ is the set of nodes and $\Ecal$ is the set of edges, denoted $(n,m)$ for an edge from node $n$ to node $m$. Note we do not assume we have a DAG in what follows and we shall comment on this further at the end of this section.

Nodes in an antichain satisfy the condition that they are not weakly connected, that is there is no directed path between the two nodes in either direction.  A directed path from $n$ to $m$ is a sequence of nodes in which consecutive nodes are linked by an edge in the correct direction \cite{N10}. That is $\{ n_j | j\in \{0,1,\ldots,\ell\}, n_0=n, n_\ell=m,    (n_j,n_{j+1}) \in \Ecal \mbox{ for } j<\ell \}$. If there is a directed path in our graph between two nodes $n,m \in \Vcal$ in \emph{either} direction i.e.\ the two nodes are \tdef{weakly connected}, we will denote this as  $n \sim m$.
An \tdef{antichain} $\Acal$ is a subset of nodes which are not weakly connected to any of the other nodes in the same antichain, that is if $n,m \in \Acal$ then $n \not\sim m$.

\begin{figure}[htb]
\centering
\includegraphics[width = 0.4\textwidth]{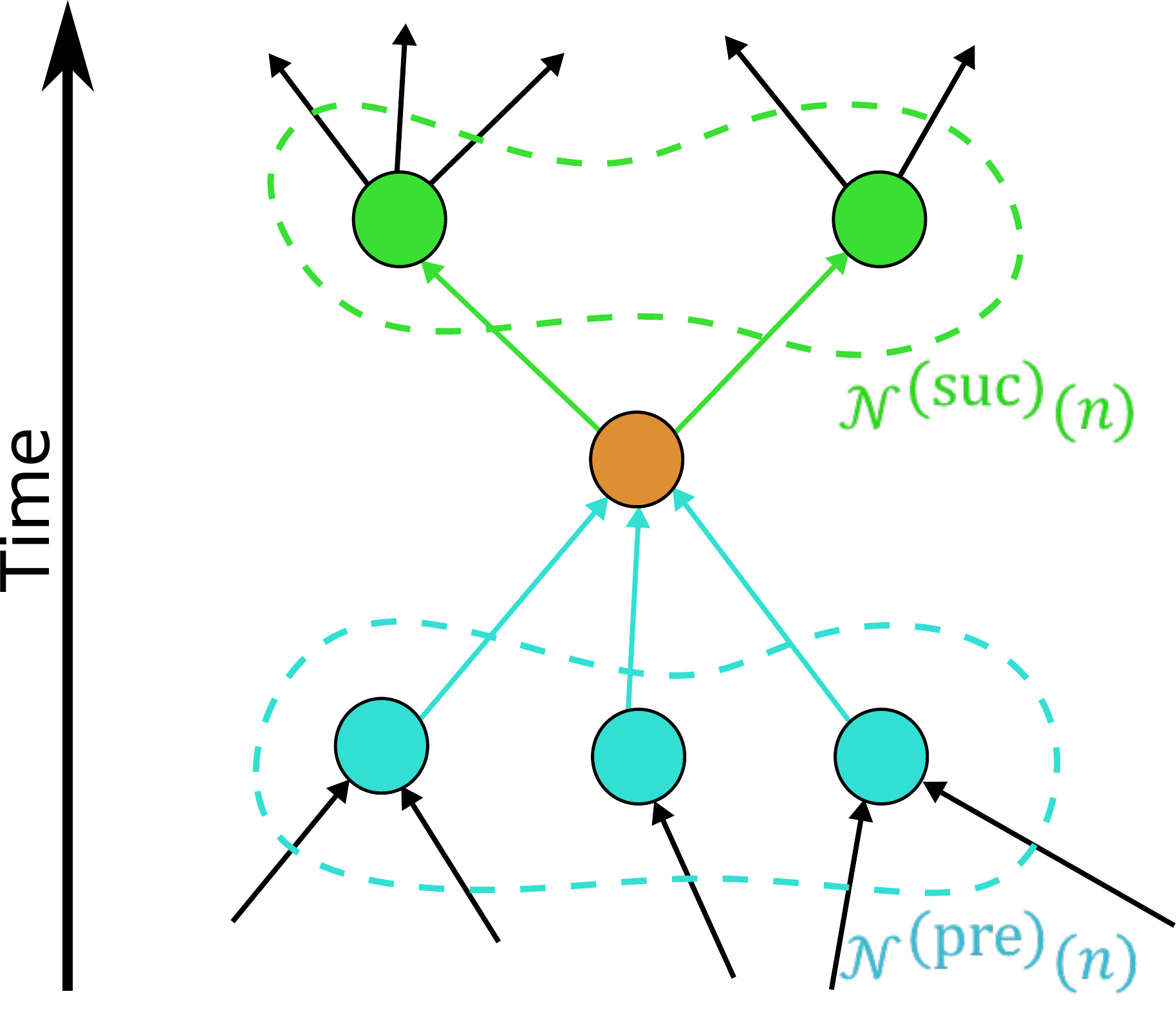}
\caption{A figure to illustrate our neighbour set definitions.}
\label{fneighbours}
\end{figure}

The second aspect is a similarity measure for two nodes, that is $\simfunc(n,m)$ is a function which increases as the nodes $n$ and $m$ become more similar.
Our aim is to use only the information encoded in the network, information which is always available. There are still many options but in our work here we will use the number of common neighbours. That is if $\Ncal(n)$ is the number of neighbours of node $n$ then we use
\beq
\simfunc(n,m) = |\Ncal(n)\cap\Ncal(m)| \, .
\label{asimfuncdef}
\eeq
In a directed graph such as our DAGs there are three natural sets of neighbours we can define.  We can use the \tdef{predecessors of $n$}, denoted  $\Ncalpre(n)$, that is the set of nodes with outgoing edges that end at $n$. Alternatively, we can use the \tdef{successors of node $n$}, denoted $\Ncalsuc(n)$, the set of nodes connected to with incoming edges that start from $n$.  That is
\beq
\Ncalpre (n) = \{ m | (m,n) \in \Ecal\} \, ,
\qquad
\Ncalsuc (n) = \{ m | (n,m) \in \Ecal\} \, ,
\label{aNcalpresuc}
\eeq
and as illustrated in \figref{fneighbours}.
Finally we can also use both sets at the same time and use $\Ncal^{\mathrm{(both)}} (n) = \Ncalpre(n) \cup \Ncalsuc(n)$ as our neighbours set.

We then need to say if a particular value for the similarity of two nodes is large or small.  To do this we define a null model, typically a randomised version of our original network, and we use the expected value in this null model for the similarity of two nodes $n$ and $m$, which we will denote as $\simfuncnull$.

\begin{figure}[htb]
\centering
\includegraphics[width = 0.95\textwidth]{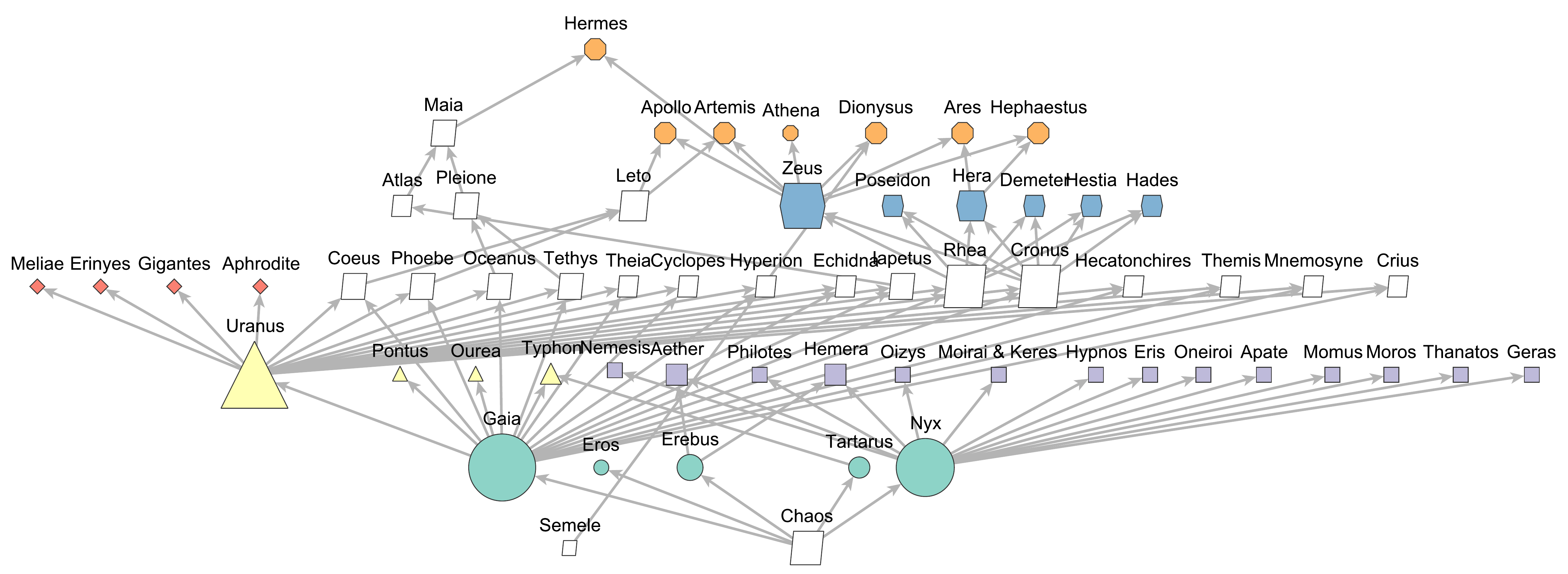}
\caption{A family tree of Greek gods based on data from \href{https://en.wikipedia.org/wiki/Family_tree_of_the_Greek_gods}{Wikipedia} \cite{WikiGreekGods} (see \secref{as:greekgods}). Links are from parents to offspring. Colours of nodes and their shapes both show grouping of deities using siblinarity based on common predecessors. For a comparison see the $\lambda=2$ example in \figref{f:greekgodslambda} which uses both successor and predecessor neighbours.
White vertices indicate nodes in an antichain community of size one. The size of a node indicates its total degree.  }
\label{fsiblings}
\end{figure}

Now we can put these elements together to define a function $S$ that measures the quality of a given partition of our network into antichains, denoted as the set $\Afrak$.  We consider a partition to be good if our antichains contain similar nodes. For instance in a family tree, we might want to group the biological siblings of mother and father pair. There is no direct biological connection between the siblings but they all have the same mother and father in common so the overlap in their precursor neighbour set, $\Ncalpre$ in the example shown in \figref{fsiblings}.
We are aiming to find a partition of our set of nodes into antichains which we refer to as an \tdef{antichain partition} and which we denote as $\Afrak$.  That is each element of the partition $\Acal \in \Afrak$ is an antichain.

Motivated by this family tree example, we call our quality function \tdef{siblinarity} and we denote this $S(\Afrak)$ for a given antichain partition $\Afrak$. The generic form we choose is
\begin{equation}\label{asiblinaritygen}
    S(\Afrak)
    =
    \sum_{\Acal\in \Afrak} \;\;
    \sum_{n \in \Acal} \;\;
    \sum_{m \in \Acal \setminus n}
    \left(\simfunc(n,m) - \simfuncnull(n,m) \right)
    \, .
\end{equation}
Note that there is no contribution from $n=m$ in this expression. This leads to the result that $S(\Afrak)$ is zero for the trivial antichain partition, the one where each nodes is in an antichain by itself, i.e.\ $\Afraktrivial$ where
\beq
 \Afraktrivial = \{ \{ v \} | v \in \Vcal \} \, .
 \label{Atrivial}
\eeq
As noted there are many possible choices for the similarity function and the null model used for comparison.
In practice what we use here is given by \eqref{asimfuncdef} which gives us
\begin{equation}\label{asiblinarity}
    S(\Afrak)
    =
    \sum_{\Acal\in \Afrak} \;\;
    \sum_{n \in \Acal} \;\;
    \sum_{m \in \Acal \setminus n}
    \left( |\Ncal(n)\cap\Ncal(m)| - \Ebb(|\Ncal(n)\cap\Ncal(m)|) \right)
    \, .
\end{equation}
The outer sum is over all antichains $\Acal$ in the antichain partition; the inner sum is over all pairs of nodes in a given antichain $\Acal$. A contribution to the total siblinarity from a pair of nodes $n$ and $m$ is equal to the size of the intersection between their neighbours (predecessors or successors or perhaps both) minus the expected value of the size of this intersection, $\Ebb(|\Ncal(n)\cap\Ncal(m)|)$.
The expected value  depends on the choice of the null model.

For instance, we use a configuration model \cite{N10} as a simple null model in which the DAG has been randomised maintaining the degree of every node and the directions of the edge but otherwise the order of the original DAG has been lost. Consider one term in our expression and so focus on a given pair of nodes $n$ and $m$ in the same antichain $\Acal$.  Then pick one of the $|\Ncal(n)|$ neighbours of node $n$, say node $p$. For simplicity we imagine that we are looking at successors so that this neighbour $p$ is at the end of an edge leaving $n$. In our simple configuration null model this neighbouring node will have in-degree $\texpect{(\kin)^2}/\texpect{\kin}$. That is neighbouring nodes of node $n$ have on average $(\texpect{(\kin)^2}-1)\kout_n/\texpect{\kin}$ incoming edges which could be at the end of edges from node $m$.  Given node $m$ has $\kout_m$ edges, the number of common neighbours may be estimated to be
\begin{equation}
 \Ebb(|\Ncal(n)\cap\Ncal(m)|)
 \approx
 \frac{(\texpect{(\kin)^2}-1)\kout_n\kout_m}{\texpect{\kin}|\Ecal|}
 \, .
\end{equation}
While this could be tried as a null model in our siblinarity expressions, we chose not to do so.  Rather we first rewrite our siblinarity in terms of matrices and then use that representation to inspire our choice of null model.


We can rewrite \eqref{asiblinarity} in terms of the adjacency matrix $\vvmatr{A}$ for our DAG \cite{SP11}. We use the convention that $\Amatr_{nm}$ is the weight of the edge from $n$ to $m$, with zero weight for no edge. Consider $\Amatrtilde$ as an effective similarity matrix obtained from the product of the adjacency matrix and its transpose.  In the case where we have an unweighted DAG, $\Amatrtilde^{\mathrm{(suc)}}=\vvmatr{A}.\vvmatr{A}^\trans$ is our successors-based similarity matrix whereas $\Amatrtilde^{\mathrm{(pre)}}=\vvmatr{A}^\trans.\vvmatr{A}$ is a similarity matrix based on predecessors, so emulating the expressions in \eqref{aNcalpresuc}. Should we choose to use both sets of neighbours then we simply use the sum of these two matrices $\Amatrtilde^{\mathrm{(both)}}=\Amatrtilde^{\mathrm{(suc)}} +\Amatrtilde^{\mathrm{(pre)}}$. Whichever of these effective similarity matrices $\Amatrtilde$ we use, it means we may write our siblinarity function in the following form:
\begin{equation}\label{asiblinaritymat}
 S({\Afrak})
  =
   \sum_{n} \;\;
   \sum_{m | m \neq n}
    \big( \tilde{A}_{nm} - \frac{\kappa_n\kappa_m}{W} \big)
   \, \delta(\Acal_n,\Acal_m)
    \, ,
    \quad
    \mbox{where } n \in \Acal_n \in \Afrak \, , \; m \in \Acal_m \in \Afrak
    \, .
\end{equation}
Here the $\delta(\Acal_n,\Acal_m)$ is one if $n$ and $m$ are in the same antichain ($\Acal_n=\Acal_m$), zero if they are in different antichains ($\Acal_n \cap \Acal_m = \emptyset$).
We define $\kappa_n$ to be the effective strength of edges attached to a node $n$ in the similarity matrix $\Amatrtilde$ so $\kappa_n = \sum_m\tilde{A}_{nm}$, while and $W = \sum_{n,m}\tilde{A}_{nm}$ is the total weight of edges in the similarity matrix.

Note that the form of \eqref{asiblinaritymat} means that we have chosen an explicit form for our null model. This matrix form \eqref{asiblinaritymat} has been chosen to emulate the modularity function \cite{NG04} used as a measure of the quality of a partition of nodes in a weighted undirected network with adjacency matrix $\Amatrtilde$. The null model we use is one in which we look at a ``second-neighbour network'' whose adjacency matrix is $\Amatrtilde$. This has the same nodes as the original DAG $\Gcal$ with undirected but weighted edges present between nodes if the are second neighbours in the original DAG.  These second neighbours in the original DAG $\Gcal$ are defined by going one step forwards and one step backwards if we are using successor neighbourhoods, and similarly for other choice of neighbourhood $\Ncal$ of \eqref{aNcalpresuc}. The null model is a configuration model \cite{N10} for this second-neighbour network.

Also note that the form given here includes non-zero diagonal entries $\tilde{A}_{nn}$ with corresponding contributions to the strength's $\kappa_n$. For instance if the original DAG $\Gcal$ is unweighted then $\tilde{A}_{nn}$ is the number of (first) neighbours of node $n$ in the DAG $\Gcal$. One could eliminate the self-loops from the second-neighbour network producing a non-backtracking form for the adjacency matrix say $\tilde{\vvmatr{A}}^{\mathrm{(NBT)}}$ instead of $\Amatrtilde$. For instance we could use $\tilde{A}^{\mathrm{(NBT,suc)}}_{mn} = (\tilde{A}^{\mathrm{(suc)}})_{mn} - \kout_n \delta_{mn}$ instead of $\Amatrtilde^{\mathrm{(suc)}}$. We see no strong reason to use this non-backtracking form and have not considered $\tilde{\vvmatr{A}}^{\mathrm{(NBT)}}$ here. Equally, apart from algebraic simplicity,  we can no reason not to use the non-backtracking form $\tilde{\vvmatr{A}}^{\mathrm{(NBT)}}$ but choose not to pursue this further here.

The big difference between siblinarity and modularity is that for our context, our partitions are restricted to be antichains, something implicit in our $\Afrak$ notation. Apart from this important restriction, we are working on the modularity of a derived weighted but undirected network with adjacency matrix given by $\Amatrtilde$. If our original DAG was unweighted, this effective adjacency matrix counts the number of `routes' (not a path in the usual precise definition used in graph theory) which consist of one forward and one backwards step on our original DAG.

\subsection{Resolution}\label{a:resolution}

It is worth noting that we can control the resolution of obtained partition by scaling the null model contribution in the siblinarity function \eqref{asiblinaritygen} by a parameter $\lambda$. This mimics one way that the resolution can be changed for community detection using modularity \cite{RB06}. In our case we suggest a modified form for the generic siblinarity function
\begin{equation}
    S(\Afrak,\lambda)
    =
    \sum_{\Acal\in \Afrak} \;\;
    \sum_{n \in \Acal} \;\;
    \sum_{m \in \Acal \setminus n}
    \left(\simfunc(n,m) - \lambda \, \simfuncnull(n,m) \right)
    \, .
\label{ae:sibmod}
\end{equation}
This clearly reduces to the original equation, \eqref{asiblinarity}, when $\lambda=1$. Large values of $\lambda$ would yield smaller antichains as adding nodes to an antichain produces a penalty. So for large enough $\lambda$, the antichain partition which maximises the modified siblinarity will be where each antichain contains just one node. When $\lambda$ is zero, any two nodes which are not connected by a path but share at least one shared neighbour will increase the siblinarity value if put in the same antichain. So for small $\lambda$ we expect the maximal modified siblinarity is likely to be something that has the fewest number of antichains. In particular, a negative $\lambda$ will allow nodes with no path between them and with no common neighbours to to have larger siblinarity values if they are in the same antichain rather than each node being in an antichain of one node.

\begin{figure}
    \centering
    \includegraphics[width= 0.8\linewidth]{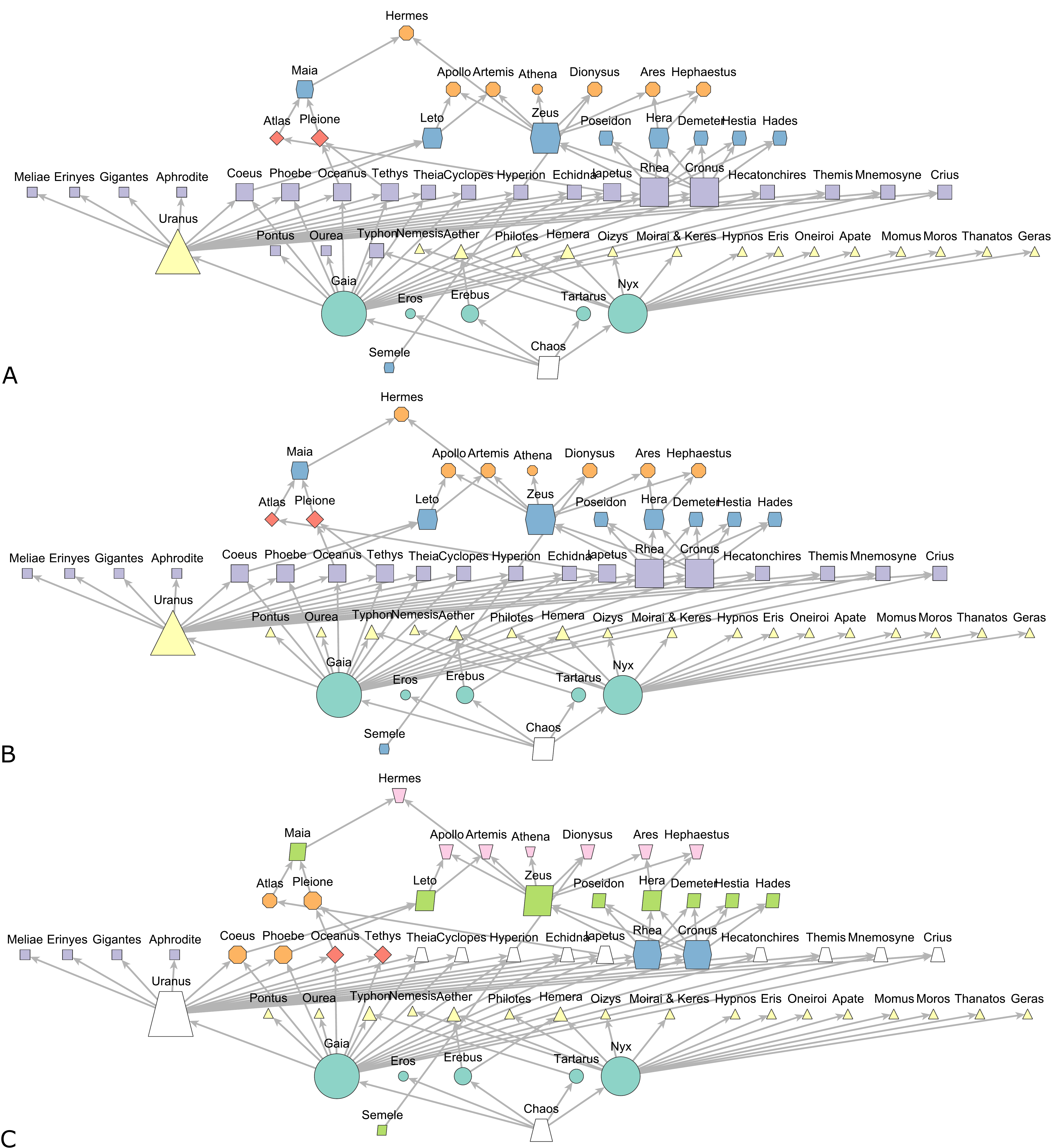}
    \caption{Antichains in the family tree of Greek Gods \cite{WikiGreekGods} (see \secref{as:greekgods}), obtained by maximising the modified siblinarity $S({\Afrak},\lambda)$ \eqref{ae:sibmodmat} using different resolution parameters $\lambda$: in A $\lambda=0.5$, in B $\lambda=1.5$, in C $\lambda=2$. Different colours an shapes show different siblinarity communities, obtained by considering both, future and past neighbours of nodes. As expected, when $\lambda$ is small, communities are large and very close to height antichains. Larger values of $\lambda$ produce more refined partitions.
    White vertices indicate nodes in an antichain community of size one.
    Note that we can compare this $\lambda=2$ example against the antichain communities in \figref{fsiblings} as the latter uses predecessors unlike here where both successor and predecessor neighbours are used to evaluate siblinarity.
    }
    \label{f:greekgodslambda}
\end{figure}

To illustrate this idea, we find antichain communities by maximising a modified form of our usual matrix siblinarity \eqref{asiblinaritymat}, namely
\begin{eqnarray}\label{ae:sibmodmat}
 S({\Afrak},\lambda)
  &=&
   \sum_{\Acal \in \Afrak} \;\;
   \sum_{n \in \Acal} \;\;
   \sum_{m \in \Acal \setminus n}
   \big( \tilde{A}_{nm} - \lambda \frac{\kappa_n\kappa_m}{W} \big)
    \, .
\end{eqnarray}
We use $S({\Afrak},\lambda)$ to find antichain communities in the family tree of the Greek gods \cite{WikiGreekGods} (see \secref{as:greekgods}) shown in \figref{f:greekgodslambda}. As expected, when $\lambda$ is small, such as in A, communities are large and very close to height antichains. Larger values of $\lambda$ in B and C produce more refined partitions. The same behaviour can be sen in a very simple example in \secref{a:example}.

\clearpage

\section{A Simple example}\label{a:example}

\begin{figure}[htb]
\centering
\begin{tabular}{c c c }
\includegraphics[height=6cm]{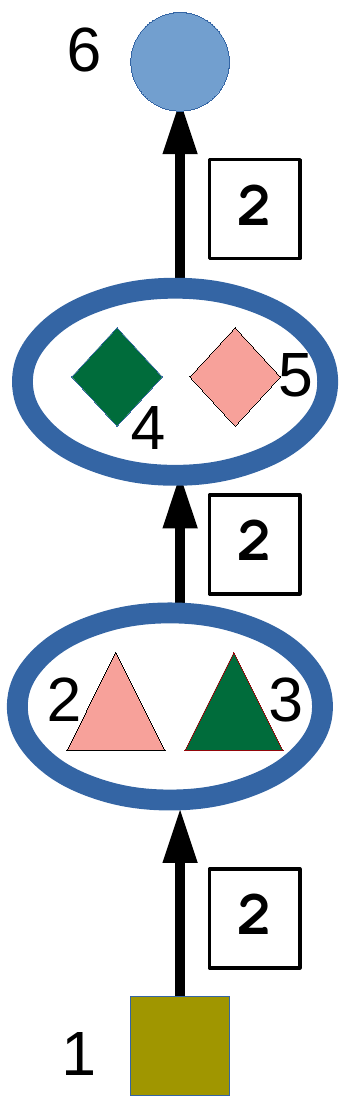} 
&
\includegraphics[height=6cm]{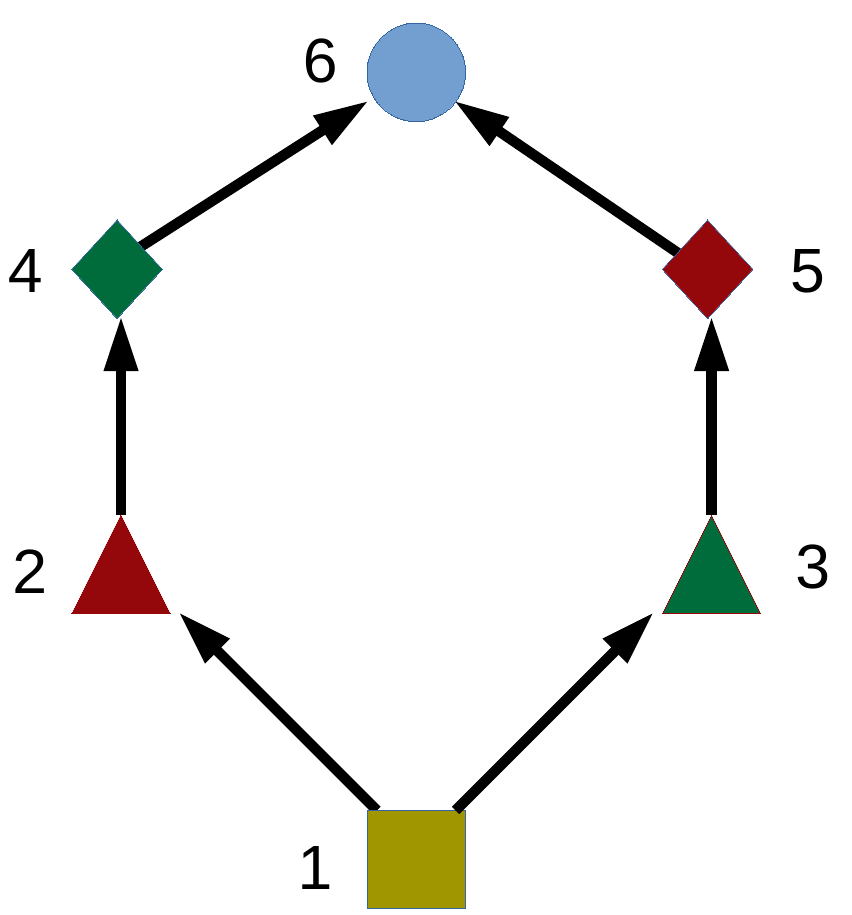}         
&
\includegraphics[height=6cm]{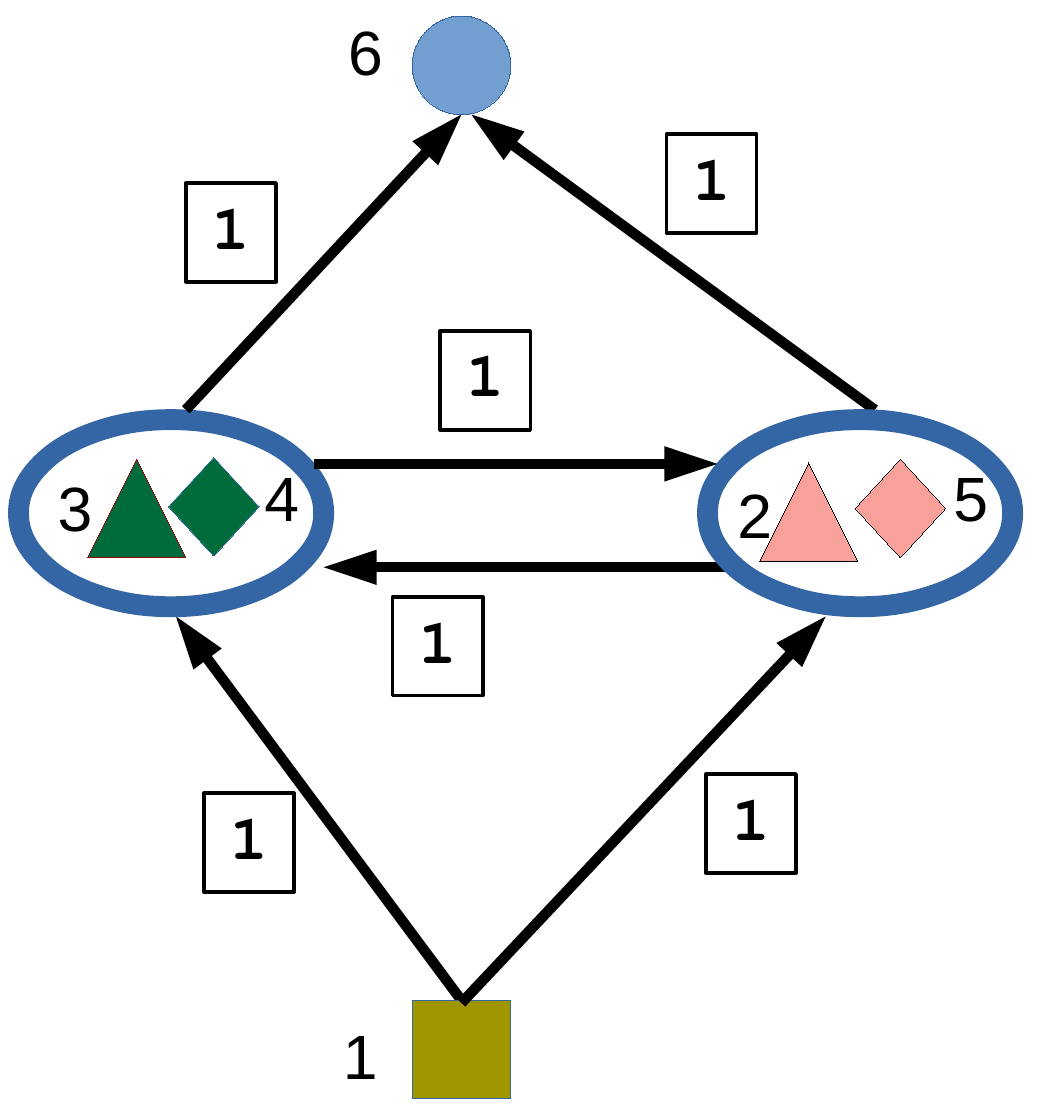} 
\\
\parbox[t]{0.3\textwidth}{\centering
 $\Hcal$ for $\Afrak_2$
 \\
 $\left\{
 \{ 1\}, \;
 \{ 2,3\}, \;
 \{ 4,5\}, \;
 \{ 6\}
 \right\}$.}
 &
 $\Gcal$
 &
\parbox[t]{0.3\textwidth}{\centering
 $\Hcal$ for
 $\Afrak_3$
 \\
 $
 \left\{
 \{ 1\}, \;
 \{ 2,5\}, \;
 \{ 3,4\}, \;
 \{ 6\}
 \right\}$.
 }
\end{tabular}
\caption{In the centre the simple example $G$ considered in the text.  On the left is the reduced network $\Gtilde_2$ based on the height antichain partition $\Afrak_2$ of \eqref{part2}.  On the right is the reduced network $\Gtilde_3$ based on the antichain partition $\Afrak_3$ of \eqref{part3}.  Note that this last antichain partition $\Afrak_3$ produces a reduced network which is a directed graph with cycles. In the networks on the left and right, the numbers in squares indicate the weight of the nearest edge.  The central network is unweighted. For the reduced networks on the left and right, the ovals are the nodes in the centre but within they show which vertices were merged to form the new node in the reduced graph.}
\label{fsmallexample}
\end{figure}

Consider the network $G$ shown in the centre of \figref{fsmallexample}.
Numbering the vertices from the bottom up we have edges $(1,2)$,  $(1,3)$,  $(2,4)$,  $(3,5)$,  $(4,6)$, and  $(5,6)$ where $(m,n)$ is an edge from vertex $m$ to vertex $n$.
Using a convention that $A_{mn}$ is an edge from $m$ to $n$ we have that columns are labelled by $n$ and rows are labelled by $m$.  So $G$ in \figref{fsmallexample} has adjacency matrix
\begin{equation}
\Amatr
 =
 \begin{pmatrix}
 0 & 1 & 1 & 0 & 0 & 0 \\
 0 & 0 & 0 & 1 & 0 & 0 \\
 0 & 0 & 0 & 0 & 1 & 0 \\
 0 & 0 & 0 & 0 & 0 & 1 \\
 0 & 0 & 0 & 0 & 0 & 1 \\
 0 & 0 & 0 & 0 & 0 & 0
 \end{pmatrix}
\end{equation}

In this example we will also use the following definition of siblinarity
\begin{eqnarray}
 \Shat({\Afrak},\lambda)
  &=&
   \sum_{\Acal \in \Afrak} \sum_{n \in \Acal} \sum_{m \in \Acal }
   \big( \tilde{A}_{nm} - \lambda \frac{\kappa_n\kappa_m}{W} \big)
    \, ,
    \label{a:sibdef}
    \\
    \nonumber
    &&
     \;\;
    \kappa_n := \sum_m\tilde{A}_{nm}
    \, ,
    \;\;
    W = \sum_{n,m}\tilde{A}_{nm}
    \, .
\end{eqnarray}
Note that we are not excluding $m=n$ in this definition of siblinarity $\Shat({\Afrak},\lambda)$
(with $\Shat({\Afrak})=\Shat({\Afrak},\lambda=1)$) unlike the definition of $S({\Afrak})$ of \eqref{ae:sibmodmat}.
The difference is a constant independent of $\Afrak$. Since $S({\Afrak})$ of \eqref{a:sibdef} is zero for the trivial antichain partition $\Afraktrivial$ of \eqref{Atrivial}\vnote{where does this refer to?}, in which every node is placed in their own element of the partition, this difference is equal to the value of $\Shat({\Afrak})$ for that trivial antichain partition. In our case this is $\Shat({\Afrak_1})$ below.
We will choose to use a successor form of the two-step matrix in \eqref{a:sibdef}, that is ${\Amatrtilde}^{\mathrm{(suc)}}=\Amatr.\Amatr^\trans$, so that
\begin{equation}
{\Amatrtilde}^{\mathrm{(suc)}}
 =
 \begin{pmatrix}
 2 & 0 & 0 & 0 & 0 & 0 \\
 0 & 1 & 0 & 0 & 0 & 0 \\
 0 & 0 & 1 & 0 & 0 & 0 \\
 0 & 0 & 0 & 1 & 1 & 0 \\
 0 & 0 & 0 & 1 & 1 & 0 \\
 0 & 0 & 0 & 0 & 0 & 0
 \end{pmatrix}
\end{equation}

\subsubsection*{Each vertex forms on antichain}

Now suppose we use this for a partition where each element of the partition is a single vertex.  This is always an antichain partition. Here we have
\begin{equation}
 \Afrak_1 =
 \left\{
 \{ 1\}, \;
 \{ 2\}, \;
 \{ 3\}, \;
 \{ 4\}, \;
 \{ 5\}, \;
 \{ 6\}
 \right\}
 \label{part1}
\end{equation}
and we find that the siblinarity value is
\begin{eqnarray}
    \Shat(\Afrak_1,\lambda)
    &=&
      \left( 2 - \lambda \frac{2.2}{8} \right)
    + \left( 1 - \lambda \frac{1.1}{8} \right)
    + \left( 1 - \lambda \frac{1.1}{8} \right)
    + \left( 1 - \lambda \frac{2.2}{8} \right)
    + \left( 1 - \lambda \frac{2.2}{8} \right)
    + 0
    \nonumber \\
    &=&
    6 - \lambda\frac{14}{8}
    \, .
    \label{sib1}
\end{eqnarray}

\subsubsection*{The Height Antichain Partition}

The height antichain partition (same as the depth antichain partition here) is
\begin{equation}
 \Afrak_2 =
 \left\{
 \{ 1\}, \;
 \{ 2, 3\}, \;
 \{ 4, 5\}, \;
 \{ 6\}
 \right\} \, .
 \label{part2}
\end{equation}
This has siblinarity equal to
\begin{equation}
    \Shat(\Afrak_2,\lambda)
    =
         \left( 2 - \lambda   \frac{2.2}{8}\right)
    +    \left( 2 - \lambda 3.\frac{1.1}{8}\right)
    +    \left( 4 - \lambda 3.\frac{2.2}{8}\right)
    + 0
    =
    6 - \lambda\frac{19}{8}
    \, .
    \label{sib2}
\end{equation}

\subsubsection*{Antichain Partition with Cyclic Derived Graph}

For an antichain partition of
\begin{equation}
 \Afrak_3 =
 \left\{
 \{ 1\}, \;
 \{ 2, 5\}, \;
 \{ 3, 4\}, \;
 \{ 6\}
 \right\}
 \label{part3}
\end{equation}
the derived graph $\Gtilde$ will be directed but not cyclic as shown in \figref{fsmallexample}.
The siblinarity for this antichain partition is
\begin{eqnarray}
    \Shat(\Afrak_3,\lambda)
    &=&
         \left( 2 - \lambda   \frac{2.2}{8}\right)
    +    \left( 2 - \lambda   [\frac{1.1}{8} + \frac{1.2}{8} + \frac{2.2}{8}]\right)
    +    \left( 2 - \lambda   [\frac{1.1}{8} + \frac{1.2}{8} + \frac{2.2}{8}]\right)
    + 0
    \nonumber \\
    &=&
    6 - \lambda\frac{18}{8}
    \, .
    \label{sib3}
\end{eqnarray}

\subsubsection*{Antichain Partition Four }

Another possible antichain partition is
\begin{equation}
 \Afrak_4 =
 \left\{
 \{ 1\}, \;
 \{ 2\}, \;
 \{ 3\}, \;
 \{ 4, 5\}, \;
 \{ 6\}
 \right\}
 \label{part4}
\end{equation}
This gives
\begin{equation}
    \Shat(\Afrak_4,\lambda)
    =
         \left( 2 - \lambda   \frac{2.2}{8}\right)
    +    \left( 1 - \lambda   \frac{1.1}{8}\right)
    +    \left( 1 - \lambda   \frac{1.1}{8}\right)
    +    \left( 3 - \lambda 3.\frac{2.2}{8}\right)
    + 0
        =
    7 - \lambda\frac{18}{8}
    \, .
    \label{sib4}
\end{equation}

\subsubsection*{Antichain Partition Five}

The fifth antichain partition we consider is
\begin{equation}
 \Afrak_5 =
 \left\{
 \{ 1\}, \;
 \{ 2,3\}, \;
 \{ 4 \}, \;
 \{ 5\}, \;
 \{ 6\}
 \right\}
 \label{part5}
\end{equation}
This is similar to antichain four but it is not related by any symmetry so gives a different result.
\begin{equation}
    \Shat(\Afrak_5,\lambda)
    =
         \left( 2 - \lambda  \frac{2.2}{8}\right)
    +    \left( 2 - \lambda 3.\frac{1.1}{8}\right)
    +    \left( 1 - \lambda  \frac{2.2}{8}\right)
    +    \left( 1 - \lambda  \frac{2.2}{8}\right)
    + 0
    =
    6 - \lambda\frac{15}{8}
    \, .
    \label{sib5}
\end{equation}

\subsubsection*{Antichain Partition Six and Seven}

The last two possible antichain partitions have the same siblinarity by symmetry
\begin{equation}
 \Afrak_6 =
 \left\{
 \{ 1\}, \;
 \{ 2,5\}, \;
 \{ 3 \}, \;
 \{ 4\}, \;
 \{ 6\}
 \right\}
 \label{part6}
\end{equation}
and
\begin{equation}
 \Afrak_7 =
 \left\{
 \{ 1\}, \;
 \{ 2\}, \;
 \{ 3,4 \}, \;
 \{ 5\}, \;
 \{ 6\}
 \right\}
 \label{part7}
\end{equation}
These have siblinarity values of
\begin{eqnarray}
    \Shat(\Afrak_6,\lambda)
    =
    \Shat(\Afrak_7,\lambda)
    &=&
         \left( 2 - \lambda   \frac{2.2}{8}\right)
    +    \left( 1 - \lambda   \frac{1.1}{8}\right)
    +    \left( 2 - \lambda   [\frac{1.1}{8} + \frac{1.2}{8} + \frac{2.2}{8} ] \right)
    \nonumber \\
    && \quad \quad
    +    \left( 1 - \lambda   \frac{2.2}{8}\right)
    + 0
    \nonumber \\
    &=&
    6 - \lambda\frac{16}{8}
    \, .
    \label{sib6}
\end{eqnarray}

\subsubsection*{Siblinarity Maximisation}

The results for all the possible antichains for the DAG $G$ of \figref{fsmallexample} are summarised in \tabref{texsibvalues}. The best partition for maximum siblinarity is not in fact the height partition $\Afrak_2$,
(this is second best) but is the fourth partition  $\Afrak_4$.
This because the two antichains $\{2\}$ and $\{3\}$ are preferred to the single antichain  $\{2,3\}$ and this is because nodes 2 and 3 have no common successors.  Had we used a siblinarity that involved predecessors in some way then $\{2,3\}$ would have been preferred.

When we introduce the resolution parameter $\lambda$ in \eqref{ae:sibmodmat}, the modified siblinarity gives the height antichain community $\Afrak_2$ \eqref{part2}, a maximal antichain partition, as the best solution for $\lambda<0$. For $0<\lambda<2$ we find the fourth antichain $\Afrak_4$ of \eqref{part4} is the optimal.  The trivial partition with each antichain containing one node, $\Afrak_1$ of \eqref{part1}, has the largest modified siblinarity for $\lambda>2$.

\renewcommand{\arraystretch}{1.5}
\begin{table}[htb]
\centering
\begin{tabular}{cl|c|r|c|r}
  \multicolumn{2}{c|}{Antichain Partition } & \multicolumn{2}{c|}{Siblinarity} & \multicolumn{2}{c}{Mod.Siblinarity} \\
   & & $\Shat(\Afrak)$ & $S(\Afrak)$   & $\Shat(\Afrak,\lambda)$ & $S(\Afrak,\lambda)$   \\ \hline\hline
  $\Afrak_1$ & $\left\{
 \{ 1\}, \;
 \{ 2\}, \;
 \{ 3\}, \;
 \{ 4\}, \;
 \{ 5\}, \;
 \{ 6\}
 \right\}$ &
  $\frac{34}{8}$ & $\frac{0}{8}$ &
  $6-\lambda\frac{14}{8}$ & $0$
  \\ \hline
  $\Afrak_2$ & $\left\{
 \{ 1\}, \;
 \{ 2,3\}, \;
 \{ 4,5\}, \;
 \{ 6\}
 \right\}$ &
  $\frac{37}{8}$ & +$\frac{3}{8}$ &
  $7-\lambda\frac{19}{8}$ & $1- \lambda\frac{5}{8}$
  \\ \hline
  $\Afrak_3$ & $\left\{
 \{ 1\}, \;
 \{ 2,5\}, \;
 \{ 3,4\}, \;
 \{ 6\}
 \right\}$ &
 $\frac{30}{8}$  & -$\frac{4}{8}$ &
  $6-\lambda\frac{18}{8}$ & $-\lambda\frac{4}{8}$
  \\ \hline
  $\Afrak_4$ & $\left\{
 \{ 1\}, \;
 \{ 2\}, \;
 \{ 3\}, \;
 \{ 4,5\}, \;
 \{ 6\}
 \right\}$ &
  $\frac{38}{8}$ & +$\frac{4}{8}$ &
  $7-\lambda\frac{18}{8}$ & $1-\lambda\frac{4}{8}$
  \\ \hline
  $\Afrak_5$ & $\left\{
 \{ 1\}, \;
 \{ 2,3\}, \;
 \{ 4\}, \;
 \{ 5\}, \;
 \{ 6\}
 \right\}$ &
  $\frac{33}{8}$  & -$\frac{1}{8}$ &
  $6-\lambda\frac{15}{8}$ & $-\lambda\frac{1}{8}$
  \\ \hline
  $\Afrak_6$ & $\left\{
 \{ 1\}, \;
 \{ 2,5\}, \;
 \{ 3\}, \;
 \{ 4\}, \;
 \{ 6\}
 \right\}$ &
  $\frac{32}{8}$ & -$\frac{2}{8}$ &
  $6-\lambda\frac{16}{8}$ & $-\lambda\frac{2}{8}$
  \\ \hline
  $\Afrak_7$ & $\left\{
 \{ 1\}, \;
 \{ 2\}, \;
 \{ 3,4\}, \;
 \{ 5\}, \;
 \{ 6\}
 \right\}$ &
  $\frac{32}{8}$ & -$\frac{4}{8}$ &
  $6-\lambda\frac{16}{8}$ &  $-\lambda\frac{2}{8}$
\end{tabular}
\caption{A table of siblinarity values for the different antichain partitions of the graph $G$ of \figref{fsmallexample}. We are using the modified versions of siblinarity with a resolution parameter $\lambda$, that is  $S(\Afrak,\lambda)$ of \eqref{ae:sibmodmat} and $\Shat(\Afrak,\lambda)$ of \eqref{a:sibdef}.  The $\lambda=1$ values are also given as $S(\Afrak)=S(\Afrak,\lambda=1)$ and $\Shat(\Afrak) = \Shat(\Afrak,\lambda=1)$.
Note that the values of siblinarity  $S(\Afrak,\lambda)$ and $\Shat(\Afrak,\lambda)$ are related through the value for the largest (trivial) partition $\Afrak_1$ since  $S(\Afrak,\lambda) =\Shat(\Afrak,\lambda) - \Shat(\Afrak_1,\lambda)$.}
\label{texsibvalues}
\end{table}

\section{Louvain Siblinarity Optimisation}\label{alouvain}

Having defined a quality function, we can look for a partition of our nodes into antichains, $\Afrak$, which maximises the siblinarity $S(\Afrak)$. This task faces the same challenges as most network community detection methods; there are many local minima and only approximate solutions can be found in a reasonable amount of computational time.
Here, we will discuss how to adapt the Louvain algorithm \cite{BGLL08} which is a widely used and successful methods to find communities in networks which maximise modularity.
Emulating the Louvain algorithm, our siblinarity maximisation method is an iterative greedy algorithm in which each iteration has two phases.

In the first phase of our algorithm, we start with an initial partition into antichains in which each node is assigned to its own antichain. At each subsequent step, we try to move a single node $n$ from its current antichain, $\Acal_a$, to another antichain $\Acal_b$, always choosing the configuration which maximises the siblinarity, even if that means leaving the antichains unchanged.  In our implementation, we visit each node in a fixed sequence. Once the sequence is exhausted, we sweep through the same sequence once again.  This process is continued until there are no more changes in the optimal antichain partition possible whatever node $n$ we choose to examine. That is when changing the antichain partition by moving just one node can not increase the siblinarity. This marks the end of the first phase. In principle, we can also stop this first phase at any point as every new antichain partition is, by definition better than the last. So in our algorithm we also stop this phase if we have completed a given number of sweeps since the second phase is almost certain to simplify the problem and so speed up subsequent iterations.

For each node $n$ we choose for a possible move, the change in siblinarity is calculated for removing $n$ from its current antichain, $\Acal_a$, and placing in a new antichain $A_b$. It is important to enforce the constraint that the node $n$ must not be connected to any existing node $m$ in the potential new antichain $\Acal_b$, i.e.\ we want $\Acal_b \cup \{n\}$ to be an antichain. In our algorithm, we further limit the choice of which new antichains $\Acal_b$ we examine.  If we are using siblinarity defined using a similarity measure using a neighbourhood set $\Ncal(n)$ for our nodes $n$, then we look for antichains $\Acal_b$ which contain at least one node $m \in \Acal_b$ which has a non-trivial similarity measure with our chosen node $n$, i.e.\ for us we require $|\Ncal(n)\cap\Ncal(m)|>0$. These potential new antichains for node $n$ are easy to find as this involves a two-step walk on the network starting from $n$.
If we use a neighbourhood based on successors, that is $\Ncalsuc(n)$, then we only look at antichains $\Acal_b$ which contain a predecessor $m$ of a successor of $n$.
We will call these \tdef{successor antichains}. If we look only at predecessors neighbourhoods and so use $\Ncalpre(n)$ for our neighbourhood sets, we shall refer to the resulting antichain partitions as \tdef{predecessor antichains}. There is only one other case we examine, and that is we also check the case where we allow $n$ to join a new antichain consisting of the node $n$ alone, i.e.\ $\Acal_b = \emptyset$.

The change in siblinarity $\Delta S$ is given by
\begin{equation}\label{edeltasiblinarityev}
\begin{split}
    \Delta S(\Acal_a, \Acal_b\rightarrow \Acal_a \backslash n, \Acal_b \cup \{n\}) =
         \sum_{m\in\Acal_b}             \left( |\Ncal(n)\cap\Ncal(m)| - \Ebb(|\Ncal(n)\cap\Ncal(m)|) \right)
    \\ - \sum_{q\in\Acal_a \backslash n}\left( |\Ncal(n)\cap\Ncal(q)| - \Ebb(|\Ncal(n)\cap\Ncal(q)|) \right) \, ,
    \\
    \mbox{provided } n \not\sim \Acal_b \, .
    \end{split}
\end{equation}
The first term is the contribution from the addition of node $n$ to the antichain $\Acal_b$, while the second term is the effect of the removal of node $n$ from its current antichain $\Acal_a$.
Note the condition that $n$ is not connected to any node in the existing antichain $\Acal_b$ which we denote as $n \not\sim \Acal_b$. This is needed to ensure $\Acal_b$ remains an antichain when $n$ is added. Computationally it requires a check if there are no paths in the network between any pair of nodes of an antichain. This can be done in several ways. For instance, an entry $A'_{nm}$ of a matrix $\Amatr'=\sum_{i=1}^\ell Amatr^i$, where $\ell$ is the length of the longest path in the network, is only zero if there is no path from $n$ to $m$. So if $A'_{nm}+A'_{mn}=0$, the nodes $n,m$ can be in the same antichain.

In the matrix notation of \eqref{asiblinaritymat}, the change in siblinarity is given by
\begin{equation}\label{edeltasiblinaritymat}
\begin{split}
    \Delta S(\Acal_a, \Acal_b\rightarrow \Acal_a \backslash n, \Acal_b \cup \{n\}) =
    \sum_{m\in \Acal_b            } \Big(\tilde{A}_{nm} -  \frac{\kappa_n\kappa_m}{W} \Big) -
    \sum_{q\in \Acal_a\backslash n} \Big(\tilde{A}_{nq} -  \frac{\kappa_n\kappa_q}{W} \Big)
    \\
    \mbox{provided } n \not\sim \Acal_b \, .
    \end{split}
\end{equation}

In the second phase we create an \textit{induced graph} $\Hcal=\{\Vcal_\mathrm{H},\Ecal_\mathrm{H}\}$ from the original graph $\Gcal$ and the antichain partition $\Afrak$ left at the end of phase one. Each node $a \in \Vcal_\mathrm{H}$ in this induced graph $\Hcal$ represents a single antichain, $\Acal_a\in \Afrak$, as given at the end of the previous phase. The edges between nodes of induced graph are given a weight equal to the sum of the weights of all the edges between the equivalent antichain nodes in the original graph $\Gcal$ of the induced graph. For instance, if there were $k_{ba}$ edges all of weight $1$ pointing from nodes in the antichain $\Acal_a$ to the antichain $\Acal_b$ at the end of phase one, there would be a directed edge $(a,b) \in \Ecal_\mathrm{H}$ in the induced graph with weight equal to $k_{ba}$ in the induced graph\footnote{Induced graph does not have to be a DAG: antichains are possible in graphs with cycles as \figref{fsmallexample} shows. By definition, an antichain is a subset of nodes such that there is no path between any of two of them in this subset. This is perfectly valid in any graph, however, in some they are more interesting than in others.}. In terms of matrices, if $H_{ba}$ is equal to the weight of the edge from node $a$ to $b$ in the adjacency matrix for the induced graph, then we have that
\beq
 H_{ba} = \sum_{m \in \Acal_a}\sum_{n \in \Acal_b} A_{nm} \, .
\eeq
\tcomment{Self-loops? Induced graph is not a DAG. Greedy and greedy greedy versions}

Once the induced graph is created, the algorithm continues by applying finding an antichain partition of the induced graph using siblinarity, starting with the phase one. The algorithm continues until there is no substantial increase in the siblinarity function \eqref{asiblinarity}.



\section{Basic Statistics on Antichains }\label{a:stats}

One way to look at antichains is to look at bipartite network $\Bcal(\Acal)$ associated with each antichain.
For each antichain $\Acal$ of a graph $\Gcal = (\Vcal,\Ecal)$, we define a bipartite network $\Bcal(\Acal)$.  The first type of vertex in the bipartite network are simply those in the antichain itself, $\Acal$.  The second type of vertex in $\Bcal(\Acal)$ is the set of all neighbours of the antichain vertices, that is
\beq
 \Ncal(\Acal) = \cup_{v \in \Acal} \Ncalsuc(v) \, .
 \label{NAdef}
\eeq
An edge of the bipartite graph $\Bcal(\Acal)$ is present if there is a corresponding connection between a given vertex in the antichain $\Acal$ to any of its neighbours, any vertex in $\Ncal(\Acal)$
\beq
 \Ecal(\Acal) =
 \{ (v,n) | n \in \Ncal(\Acal), \, v \in \Acal, \, (v,n) \in \Ecal \}
 \cup
 \{ (n,v) | n \in \Ncal(\Acal), \, v \in \Acal, \, (v,n) \in \Ecal \}  \, .
\eeq
In practice, we also remove loosely connected neighbour nodes but for simplicity we will not indicate that in our definitions here.
This gives us $\Bcal(\Acal) = (\Acal \cup \Ncal(\Acal), \Ecal(\Acal) )$.

There are many possible network measurements we can make on each bipartite graph which can help us understand the nature of the antichains found in any example. In our work we focussed on some of the simplest measures.
\begin{itemize}
  \item $|\Acal|$ --- The size of an antichain, i.e.\ the number of nodes in an antichain.
  \item $|\Ncal|$ --- The number of neighbours of an antichain.
  \item $\texpect{k}$ --- The average number of neighbours of nodes in an antichain.
  \beq
   \texpect{k}_{\Acal} = \frac{|\Ecal(\Acal)|}{|\Acal|}
  \eeq
  \item $\sigma(k)_{\Acal}$ --- The standard deviation of the number of neighbours of nodes in an antichain.
  \beq
   \sigma(k)_{\Acal} =  \sqrt{\frac{\sum_{n\in\Acal}(k_n-\texpect{k}_{\Acal})}{|\Acal|-1}}
  \eeq
  \item $\texpect{k}/|\Ncal| = \frac{|\Ecal(\Acal)|}{|\Acal||\Ncal|}$ --- This is the density of the bipartite graph, the number of edges divided by the maximum number possible.
\end{itemize}
The average degree and standard deviation of degree of nodes in the antichain, $\langle k\rangle$ and $\sigma(k)$, give us a picture of how the even the connections between antichain nodes and their neighbours are. If $\langle k\rangle \approx |\Ncal|$ and $\sigma(k)\approx 0$, we know that nodes of very similar degrees are joined together in an antichain, and their neighbourhoods are largely overlapping.

The ratio between  $\langle k\rangle$ and $|\Ncal|$ tell us how similar our bipartite graph is to a sparse ``zig-zag'' pattern. If  $\langle k\rangle$ is small in comparison to  $\Ncal$, then we can expect small overlap overall between all nodes in the antichain. This statistic approaches 1 if every node in the antichain is connected to every neighbour.

\section{Additional Examples of Siblinarity and Data}\label{a:extraexamples}

\subsection{Greek Gods}\label{as:greekgods}

We have named the function used to find antichain partitions ``siblinarity'' using the analogy with a family tree.  We use the number of common neighbours to link nodes in an antichain just as siblings share biological parents.  Since family trees based on biological parentage are predominantly trees, directed acyclic graphs but which have few if any loops in the undirected version, we have looked to fiction to provide a more interesting example of a family tree. We have taken \href{https://en.wikipedia.org/wiki/Family_tree_of_the_Greek_gods}{information on Greek gods from Wikipedia} \cite{WikiGreekGods} with each node representing a Greek god with edges from a god to a child. \Figref{fsiblings} shows the results of applying our siblinarity clustering siblinarity based on common predecessors to this data set. See \cite{figsharerepo} for the original data.

\subsection{Python Dependencies}

Many software programmes can be extended by adding packages, extra programmes which extend the functionality of the core package. As the number of packages grows, some of the added packages start to use the functionality of some of the other packages added to the original core application. In order to ensure that each package can run correctly, there is usually a system to noting the dependencies of each package, that is, what other packages are required in order to run any one package.

\begin{figure}[htb]
\centering
\includegraphics[width = 0.95\textwidth]{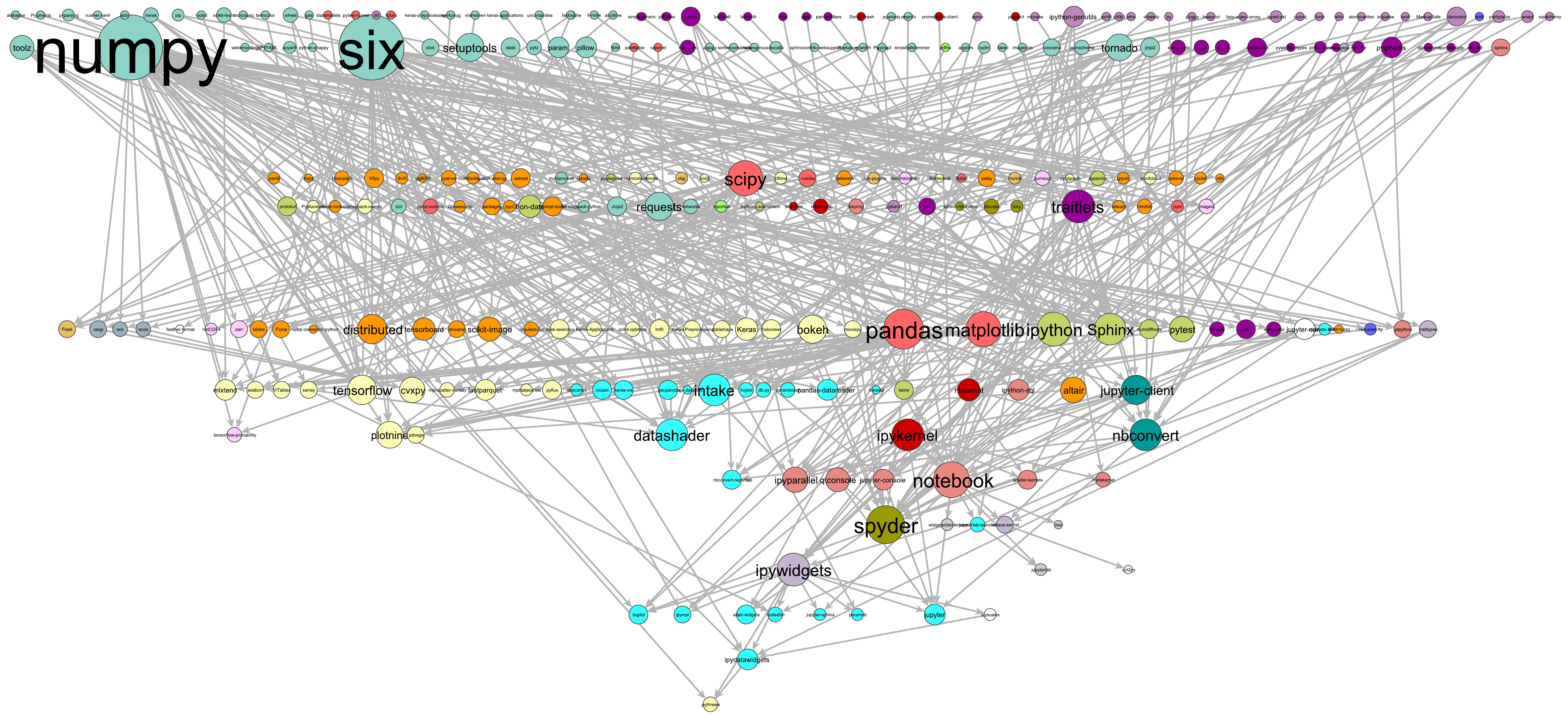}
\caption{The dependencies of \href{https://www.python.org/}{\texttt{python}} packages on one of the author's machines in 2019. Each node is a package.  Links are to the dependent package and are from the parent package needed for the siblings to function, no transitive reduction has been performed. Colours indicate the different antichain communities found using siblinarity based on both successors and predecessors.  White indicates a node in a cluster by itself. The size of the node and the size of the labels is related to the degree of each node. The vertical positioning is given by the depth of a node with small variations in the top two depths to improve visibility. An electronic version of this file is available on \cite{figsharerepo} which will allow readers to see the names of packages with small labels.}
\label{f:pydep}
\end{figure}

A good example of such an software `ecosystem' is the computer language \href{https://www.python.org/}{\texttt{python}} \cite{python}. The \href{https://pypi.org/}{\texttt{PyPI}} repository \cite{PyPI} of packages for python records over 180,000 projects at the time of writing. To illustrate the principle, we used a python \texttt{python} installation on one of the author's machines in 2019. We used V.Naik's \href{https://github.com/naiquevin/pipdeptree}{\texttt{pipdeptree}} \cite{pipdeptree} (itself a python package listed on \texttt{PyPI}) to extract the directed acyclic graph representing the package dependency DAG.  Each node is a package and we add a link from the parent package to the sibling, where the sibling package is requires the installation of the parent in order to work. The results are shown in \figref{f:pydep} and the data used is provide on \cite{figsharerepo}.

This illustrates several of the points made elsewhere.  For instance, the \texttt{scipy}, \texttt{pandas}, and \texttt{matplotlib} packages, bright red nodes in the second and third layer from the top in \figref{f:pydep}, are in the same antichain cluster.  These are often used for scientific analysis but they provide different functionality: scientific functions, data handling and plotting respectively.
They share many predecessors, such as \texttt{numpy}, since they all need to handle large quantities of numerical values efficiently. It is interesting to see that these three packages are not all at the same height so a height based clustering would not put them together.  On the other hand \texttt{sphinx} (a tool for producing documentation seen on the right of \figref{f:pydep}, coloured in bright green) is at the same height as \texttt{pandas}, and \texttt{matplotlib} (bright red nodes in \figref{f:pydep}, next to \texttt{sphinx}) but is placed in a different cluster as the packages share so few common predecessors.


\section{Price Model with Subject Fields}\label{aprice}

The Price model for citation networks \cite{P76} produces a DAG with a fat-tailed (power-law) distribution for the out-degree of nodes in our conventions which represents the citation count of papers. We modify the Price model by assigning each paper to a `field' and the edges, citations between papers, are biased so they are usually between papers in the same field. While an unrealistic model of citation networks in many ways, it contains three key features of real citation networks: the order of papers imposed by time, the fat-tailed citation count distribution, and the preference of most papers to cite papers within a similar field.  We use it as a DAG with a planted partition to enable us to make controlled comparisons between the different community detection approaches discussed.

The model defines a sequence of networks $\Gcal(t)$ where $t$ is a positive integer playing the role of time and which gives us an order to the nodes in our networks. Each graph $\Gcal(t)$ has $t$ nodes with vertex set $\Vcal(t)$ and edge set $\Ecal(t)$. In our notation, the node $u(s)$ is always the node added at step $s$ in the process, so $u(s) \in \Vcal(t)$ provided $0<s\leq t$.

The nodes in these networks are also partitioned into different fields, that is each node $u(t)$ is in one of $F$ fields.  The fields will be labelled by integers between $0$ and $(F-1)$ with $f(t) \in \{ 0, 1 \ldots, (F-1)$ denoting the field of node $u(t)$.  This creates a sequence of partitions $\Ffrak(t)$ of our nodes where $\Ffrak(t) = \{ \Fcal_f(t) | f \in \{ 0,1,\ldots,(F-1)\} \}$ and $\Fcal_f(t) \subseteq \Vcal(t)$.  A node $u(s)$, for $0<s\leq t$, belongs to a element $\Fcal_{f(s)}(t) \in \Ffrak(t)$, the set of papers at time $t$ in the same field $f(s)$ as the paper published at time $s$.


To create the next graph in the sequence, $\Gcal(t+1)$, we first add a new node $u(t+1)$ to the vertex set, so $\Vcal(t+1) = \Vcal(t)\cup\{u_{t+1}\}$.
This new node is assigned to $f$ chosen uniformly at random from the set of $F$ possible field labels.

We now add $m$ directed edges to this new node $v(t+1)$ from existing nodes $u(s)$ where $s<t$. To encode the ``cumulative advantage'' principle of Price, that is the higher the current citation count of a paper the more likely it is to be cited, we can chose nodes $u(s)$ from the existing nodes $\Vcal(t)$ in the network $\Gcal(t)$ with probability $\Pi^\mathrm{(CA)}(t,s)$ defined as\footnote{Other forms linear in $\kout(t,s)$ are also easy to work with (for example see \cite{N10}), but these variations are not our focus here. We chose to follow the same form as used in Price's original paper.}
\begin{eqnarray}
 \Pi^\mathrm{(CA)}(t,s)
 &=&
 \frac{\kout(t,s)+1}{ |\Vcal(t)| + |\Ecal(t)| } \, .
 \label{Picadef}
\end{eqnarray}
Here $\kout(t,s)$ is the number of outgoing edges from node $u(s)$ in $\Gcal(t)$, the network at time $t$.  In our conventions, these edges represent citations from later papers to the paper published at time $s$.
The planted partition representing the fields is used on top of the cumulative advantage in $\Pi^\mathrm{(CA)}(t,s)$ by ensuring that a fraction $\phi$ of the edges are chosen to lie between nodes in the same field.  So the overall probability for choosing an existing  node $u(s)$ as the source of an edge to new node $v(t)$ is, to a good approximation\footnote{For instance, we have assumed that the fraction of nodes in any one field is always $1/F$ and that the degree distribution is the same for all fields at all times.  These are good approximations at later times.},
\begin{eqnarray}
 \Pi(t,s)
 & \approx &
 \left(
 \phi F \delta(f(s),f(t+1))
 + (1-\phi) \frac{F}{(F-1)} (1-\delta(f(s),f(t+1)))
 \right)
 \Pi^\mathrm{(CA)}(t,s)
 \, .
 \label{Pidef}
\end{eqnarray}

Note that we also impose the constraint that there is at most only one edge between any two nodes and we choose the initial graph to be a transitively complete graph of size $m+1$. Neither of these constraints will have any significant effect on the measurements we make for the large networks we use in our work here.


\section{Performance of the algorithm}\label{aperformace}

We studied several metrics of the computational performance of our algorithm. Our current \texttt{Python} implementation can be found alongside the data in the Figshare repository~\cite{figsharerepo}.
In particular, we looked at the time and memory requirements of the algorithm.

\begin{figure}[htb]
\centering
\includegraphics[width = 0.5\textwidth]{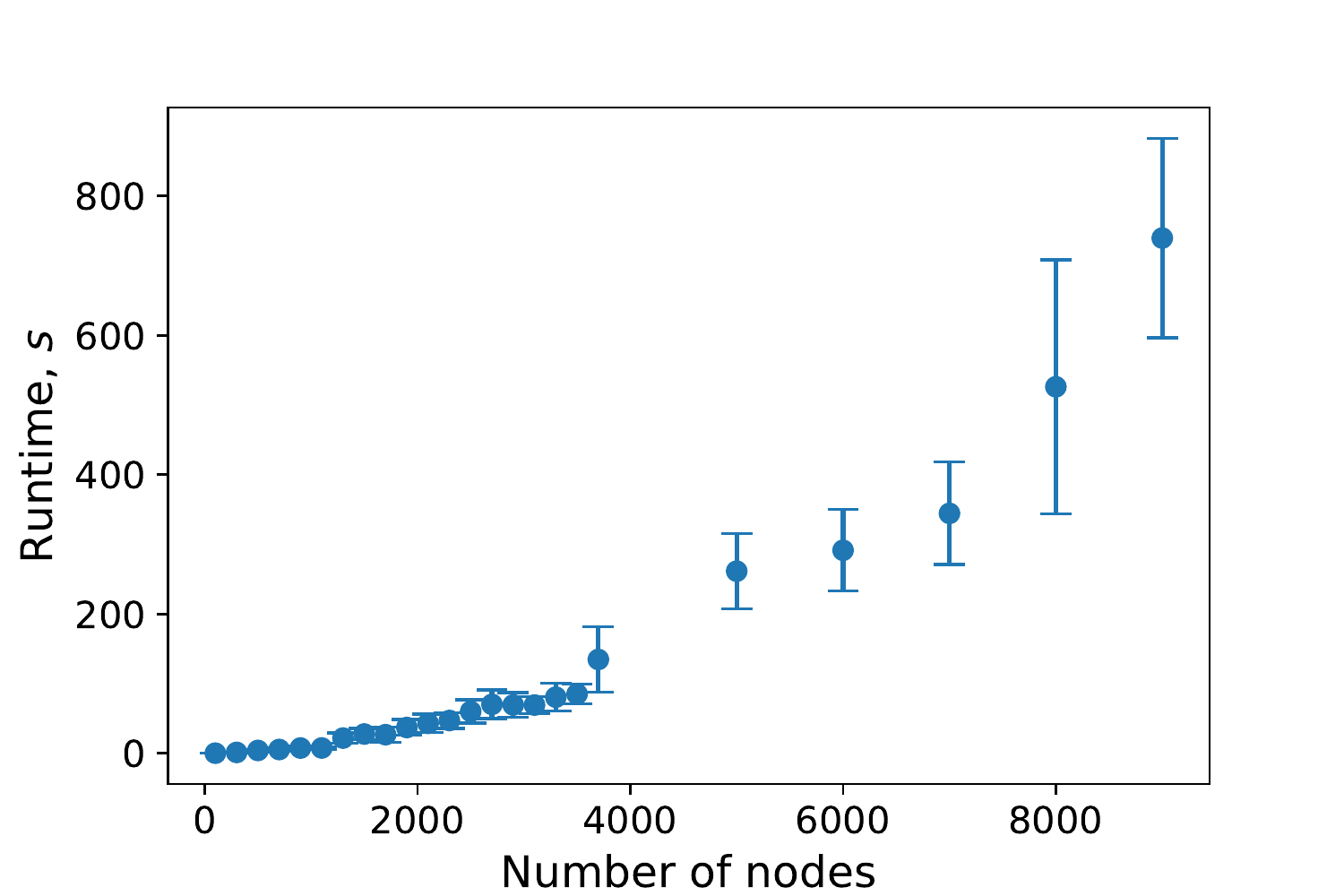}
\caption{Runtime of an algorithm, for a given size of a network, where the input network is the Price model with fields. In all networks, each node attaches $5$ edges to older nodes and 8 out of 10 times a node chooses to connect to another node in its own field. There are three fields in total. For each value of the number of nodes, $10$ networks were created.}
\label{ftime_complex}
\end{figure}

While the runtime depends on the topology of the particular network, we found that our implementation worked for graphs composed of thousands of nodes in a feasible timeframe on a standard desktop computer. In \figref{ftime_complex} we show the scaling of time consumption as the input graph size is increased. Here the networks were created using the Price model with subject fields, described in Section 3.2. The longest time to run the code for a network of 9,000 nodes was $1003s$, however, we found the variation in the runtime increases as the input graph becomes larger.

To estimate the time and memory needed to find siblinarity communities, we recognise there are two aspects to the problem.  The first aspect is that we are using a Louvain type algorithm, a type of greedy optimisation, to optimise siblinarity.  This appears to run in $O(N\ln(N))$ for a network of $N$ nodes but we have been unable to find a published study to back this up. Given every neighbour of every node is checked, it seems more likely to be $O(E\ln(N))$ for a network of $N$ nodes and $E$ edges. In our case, we use our two-step matrix $\tilde{A}$ which has the same number of nodes but far more edges $\tilde{E}$ than the DAG so the edges will scale as $\tilde{E} \sim \texpect{\kout}\texpect{\kin}N$ in terms of the average in- and out-degree of the DAG of $N$ nodes.
In terms of memory, we need to store the information in the similarity matrix $\tilde{A}$ which will scale with the number of edges stored for sparse cases, $O(\tilde{E}) \approx O((\texpect{\kout})^2N))$ (as $\texpect{\kout}=\texpect{\kin}$) with an upper bound of $O(N^2)$ for dense matrices.

The largest difference between the siblinarity optimisation code and the modularity optimisation is that we need to do an extra check for the weak connectivity of two nodes $m$ and $n$ before placing them into the same siblinarity community. In practice, we found that performing this check is feasible by directly checking the $\texttt{networkx}$ graph object each time, storing no extra information but require small searches to be performed at each step giving us a time penalty.

In theory we could impose the path constraint needed for antichains in another way and we will use this to provide an estimate for the time requirement of our method. In order to record the information about node connectivity for a network with adjacency matrix $\Amatr$, we can define a new matrix $\vvmatr{P}$ where
\beq
\vvmatr{P} = \sum_{\alpha=1}^{\ell_{\textrm{max}}}[\Amatr]^\alpha \, .
\eeq
Here $P_{mn}$ is the number of paths of any length from node $n$ to mode $m$. Here $\ell_{\textrm{max}}$ is the longest path length in the DAG (always finite for a finite DAG and often $O(\ln(N))$). In practice we only need to know if entries are zero or not as only if $P_{mn}=0$ and $P_{nm}=0$, can $n$ and $m$ be placed in the same antichain. If we use this approach and define matrix $\vvmatr{P}$ then the memory requirements will always be $O(N^2)$ as this is a dense matrix.  In terms of time, this extra step to find $\vvmatr{P}$ would require  $\ell_{\textrm{max}}$ matrix multiplications if we used matrix multiplication.  With upper triangular matrices, this could be fast.  However, given that this is a DAG, finding one path (not all paths) from one node to any others, all that is required here, and that will take $O(N+E)$ using a breadth or depth first search in a DAG of $N$ nodes and $E$ edges.  So at worst finding the zeros in matrix $\vvmatr{P}$, that is finding which pairs of nodes are connected, will take $O((1+\texpect{\kout})N^2)$ in time.

Putting this together, it appears that the memory requirements will always scale as $O(N^2)$ but the time requirements have an lower bound of $O(\texpect{\kout}N \ln(N))$ and an upper bound of $O(\texpect{\kout} N^2)$.



\section{Resolution}

As siblinarity optimisation is closely related to modularity, we should consider issues which arise when partitioning a network using modularity optimisation.
In particular, to use siblinarity appropriately one must assess which resolutions reveal meaningful siblinarity communities. One way to achieve this in practice is to study the stability of a partition at a given scale (resolution), something studied extensively for algorithms based on modularity optimisation, for example see \cite{F09,SDYB12,LDB14}, but this is an issue common to any data clustering method.

\begin{figure}[b]
\centering
\includegraphics[width = 0.45\textwidth]{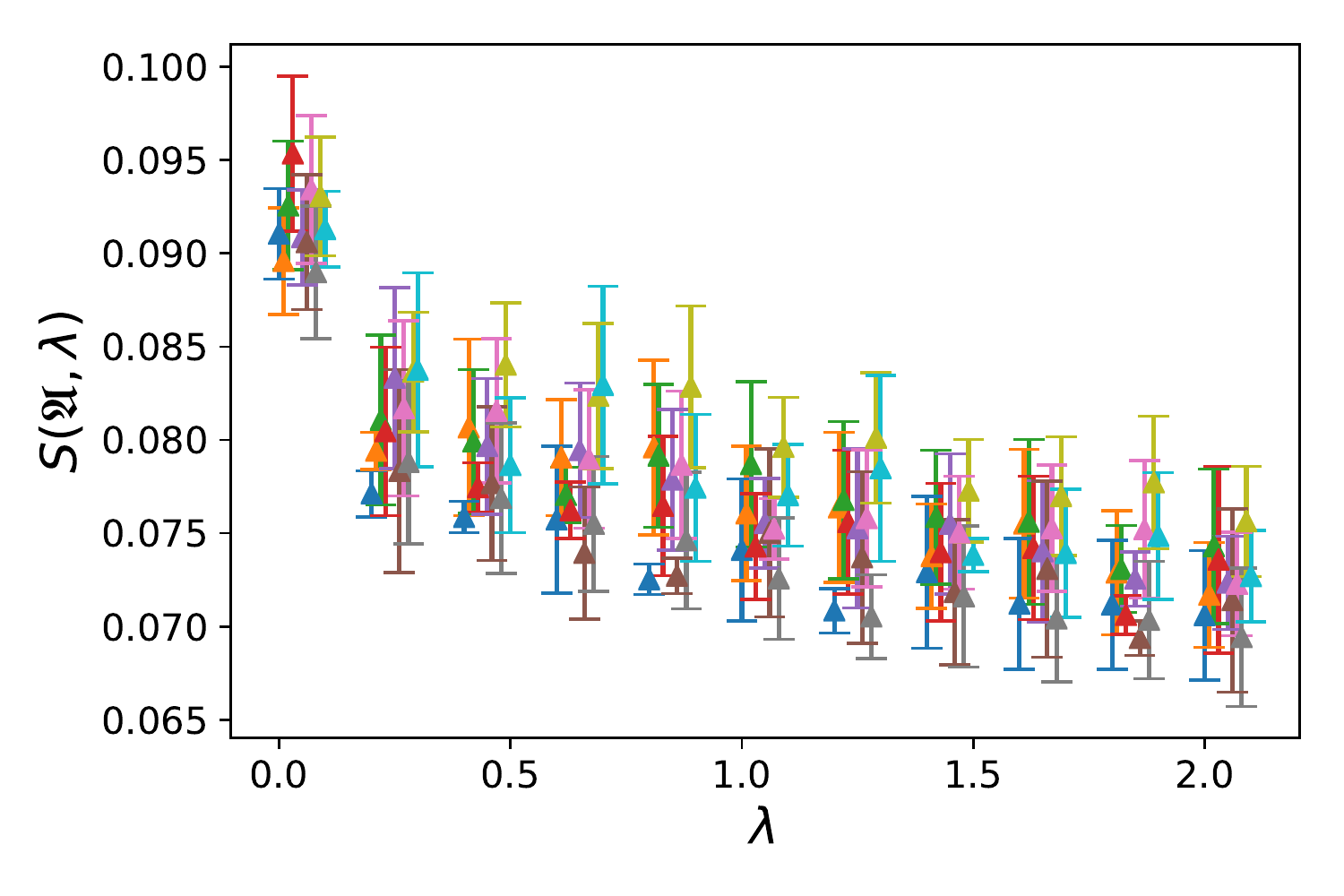}
\includegraphics[width = 0.45\textwidth]{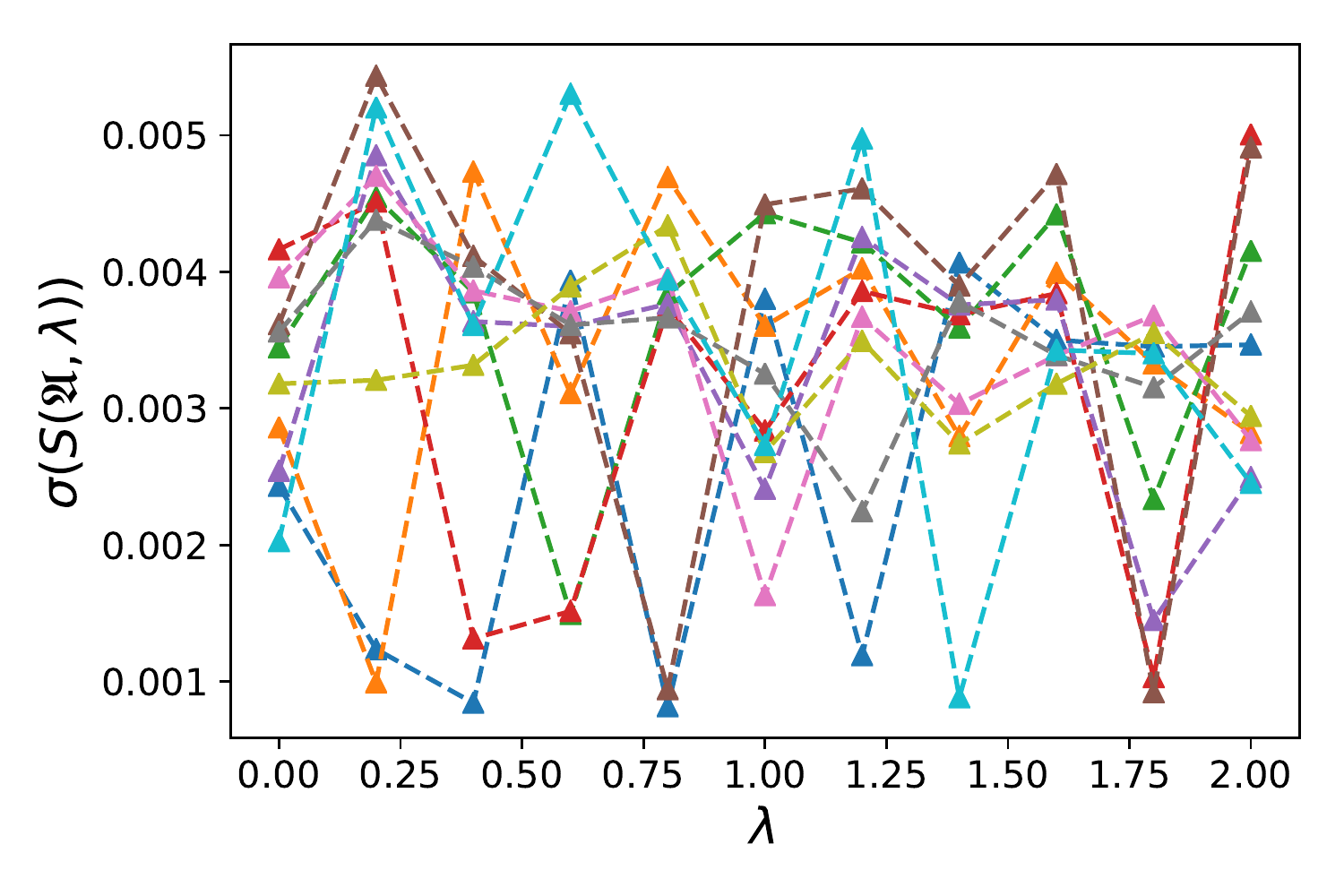} \\
\includegraphics[width = 0.45\textwidth]{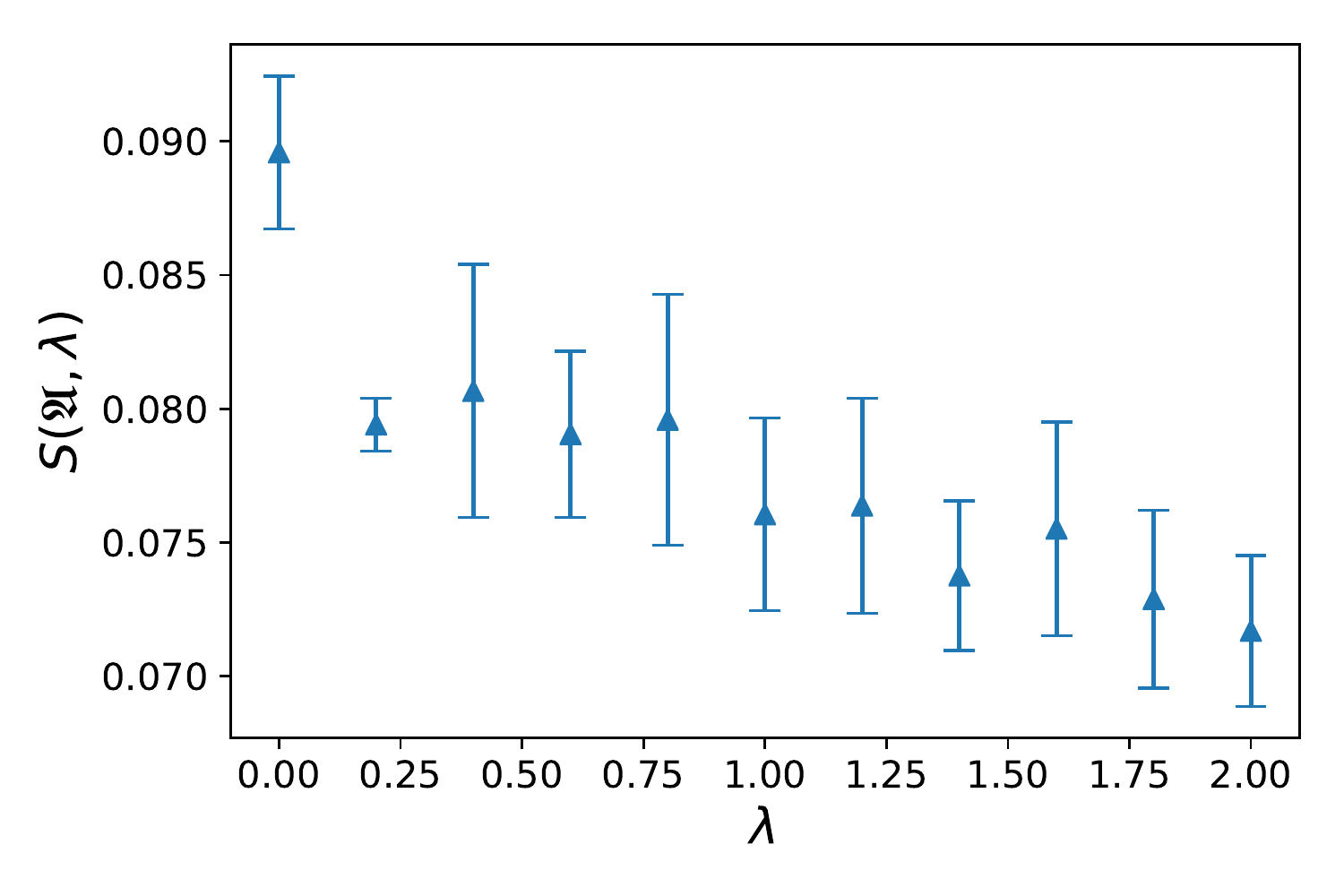}
\includegraphics[width = 0.45\textwidth]{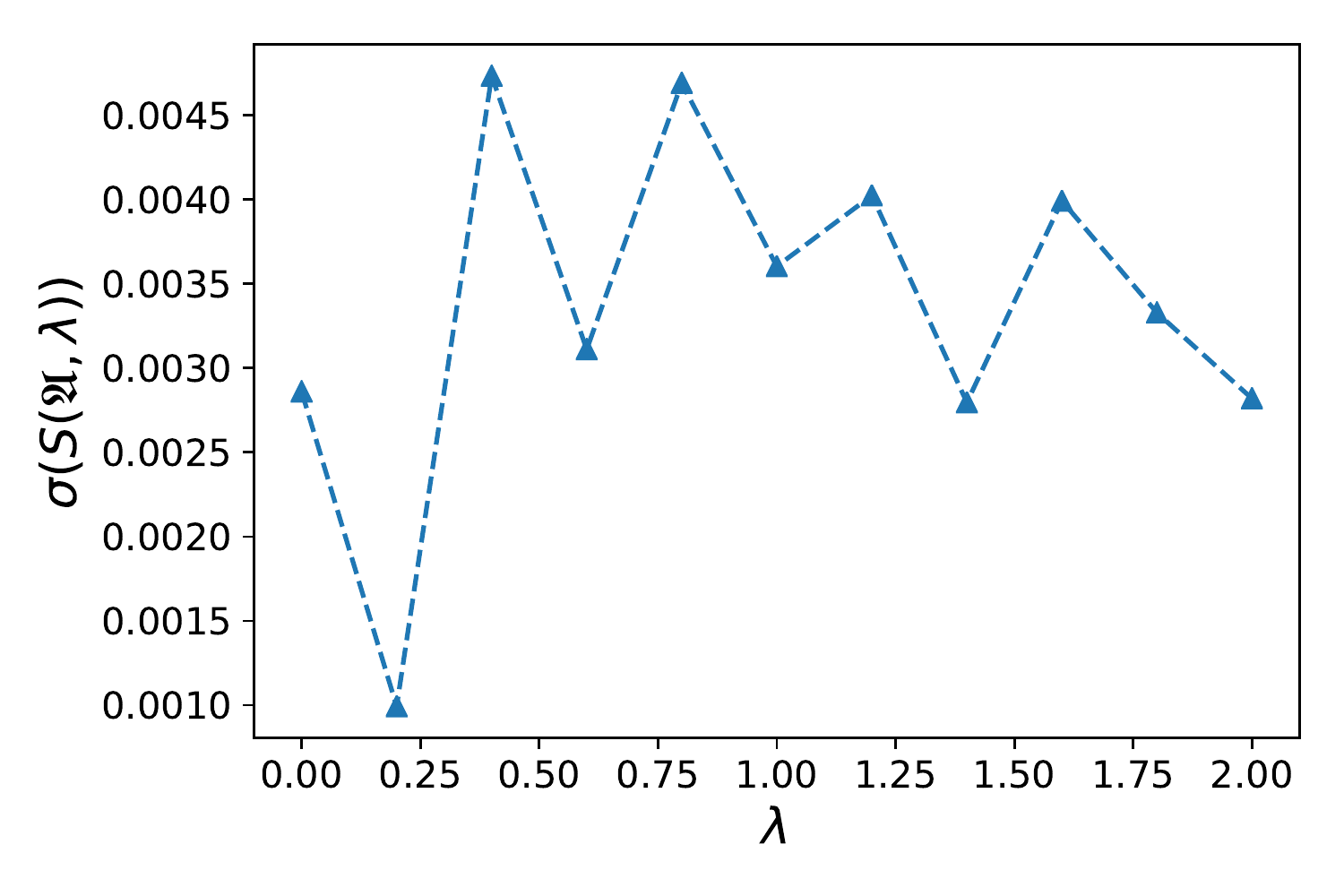}
\caption{Variations in siblinarity scores $S(\Afrak)$ of \eqref{ae:sibmodmat}. On the left, $S(\Afrak)$ for a given network is shown for a variety of $\lambda$ for multiple networks (top) and a single network (bottom). On the right, the standard deviation of $S(\Afrak)$ obtained by running the algorithm ten times for each $\lambda$ is shown for multiple networks (top) and a single network (bottom). The results from each network in the top figures are indicated using one colour and at a slightly shifted value of lambda to aid visualisation. The mean and the standard deviation were obtained by running the code 10 times for a given $\lambda$. Each network was created the Price model with subject fields, described in Section 3.2 and has $N=1000$ nodes. Each node attaches $5$ edges to older nodes. $80\%$ of the time a node chooses another node from its field (in total, there are $3$ fields). For the plot of standard deviation the dashed lines are only shown to guide the eye.}
\label{f_anysweep_vs_singlesweep_siblinarity}
\end{figure}

To demonstrate how this stability analysis can be performed for our method, we looked at the siblinarity values $S(\Afrak)$ obtained by our implementation of the siblinarity optimisation for a given network across ten different runs for different numbers of communities, here different values of our resolution parameter $\lambda$ in \eqref{ae:sibmodmat}.  As \figref{f_anysweep_vs_singlesweep_siblinarity} shows, the siblinarity scores tend to decrease with an increase of $\lambda$, as expected. However, for a given $\lambda$, the scores obtained are very similar (the same colour indicates the same random network). Lastly, we saw a minute variation in the $S(\Afrak)$ values, with standard deviation reaching the maximum of $0.005$.

\end{document}